\documentclass[pdflatex,sn-mathphys-num]{sn-jnl}

\usepackage{graphicx}%
\usepackage{multirow}%
\usepackage{amsmath,amssymb,amsfonts}%
\usepackage{amsthm}%
\usepackage{mathrsfs}%
\usepackage[title]{appendix}%
\usepackage{xcolor}%
\usepackage{textcomp}%
\usepackage{manyfoot}%
\usepackage{booktabs}%
\usepackage{algorithm}%
\usepackage{algorithmicx}%
\usepackage{algpseudocode}%
\usepackage{listings}%
\usepackage{nicefrac}

\theoremstyle{thmstyleone}%
\theoremstyle{thmstyletwo}%

\theoremstyle{thmstylethree}%
\newcommand{\kms}{\ensuremath{\, \mathrm{km}\, {\mathrm s}^{-1}}}
\newcommand{\cms}{\ensuremath{\, \mathrm{cm}\, {\mathrm s}^{-2}}}

\newcommand{\red}[1]{\textcolor{red}{#1}}
\def\msol{M$_{\odot}$}
\def\rsol{R$_{\odot}$}

\begin{document}

\title{Evidence for the emergence of stellar magnetic fields during rapid mass transfer}

\author[1,2]{Gregg A. Wade}
\author[3]{Tomer Shenar}
\author[4]{Pablo Marchant}
\author[5,6]{Michael Abdul-Masih}
\author[7,8]{Matteo Cantiello} 
\author[9]{Jason Grunhut}
\author[10]{Oleg Kochukhov}
\author[11]{Hugues Sana}
\author[12]{Selma E. de Mink}

\affil[1,*]{Dept. of Physics, Engineering Physics, and Astronomy, Queen's University Kingston, ON K7L 3N6, Canada}
\affil[2]{Dept. of Physics and Space Science, Royal Military College of Canada, Kingston, ON K7K 7B4, Canada}
\affil[3]{The School of Physics and Astronomy, Tel Aviv University, Tel Aviv, 6997801, Israel}
\affil[4]{Department of Physics and Astronomy, Proeftuinstraat 86, N3, B-9000 Ghent, Belgium}
\affil[5]{Instituto de Astrofísica de Canarias, C. Vía Láctea, s/n, 38205, La Laguna, Spain}
\affil[6]{Universidad de La Laguna, Dept. de Astrofísica, Av. Astrofísico Francisco Sánchez s/n, La Laguna, 8206, Spain}
\affil[7]{Center for Computational Astrophysics, Flatiron Institute, New York, NY, 10010, USA}
\affil[8]{Department of Astrophysical Sciences, Princeton University, Princeton, NJ, 08544, USA}
\affil[9]{Telus, 25 York St, Toronto, ON M5J 2V5, Canada}
\affil[10]{Department of Physics and Astronomy, Uppsala University, Box {524}, SE-75120 Uppsala, Sweden}
\affil[11]{Institute of Astronomy, KU Leuven, Celestijnenlaan 200D, 3001, Leuven, Belgium; Leuven Gravity Institute, KU Leuven, Celestijnenlaan 200D, box 2415, 3001, Leuven, Belgium}
\affil[12]{Max Planck Institute for Astrophysics, Karl-Schwarzschild-Straße 1, 85748 Garching bei München, Germany}

\affil[*]{wade.gregg@queensu.ca}

\abstract{
Binary interaction fundamentally alters the evolution of massive stars, yet the physical consequences of mass transfer remain poorly constrained. Here we present high-resolution spectroscopic observations of Plaskett’s Star (HD 47129) that reveal small radial-velocity variations of its magnetic secondary, coherent with the system’s 14.4-d orbital period. These measurements support an evolved, post-mass-transfer configuration in which the narrow-lined primary is a partially Roche-lobe-filling stripped star with a mass of $5.9^{+1.8}_{-1.4},M_\odot$, while the secondary is a $40.8^{+9.3}_{-6.7},M_\odot$, rapidly rotating magnetic accretor. The system provides a rare opportunity to observe the immediate aftermath of mass transfer in a massive binary and offers compelling evidence linking binary interaction to the emergence of strong stellar magnetism. Because magnetic fields regulate angular-momentum loss and transport, their formation through mass transfer could alter the subsequent evolution of massive binaries, with consequences for stellar populations, supernova progenitors and yields, and the formation of gravitational-wave sources.
}


\maketitle
%
%



\section*{Introduction}



Massive stars - those stars beginning their lives with masses greater than about 8 times that of the Sun - represent the radiative and mechanical powerhouses of galaxies. Their intense and energetic radiation fields and dense, supersonic stellar winds sculpt the structure of their host galaxies, and enrich those galaxies with their nucleosynthetic products upon their deaths as supernovae. Understanding the formation and evolution of massive stars is therefore critical to understanding the evolution of galaxies and the Universe.

Unlike lower-mass stars, the large majority of massive stars are known to form in binary or higher-order multiple systems \cite{1998A&ARv...9...63V,2007ApJ...670..747K,2012Sci...337..444S}. As these stars evolve, they expand to fill their Roche lobes (the regions of gravitational influence surrounding stars in a binary system defined by a gravito-centrifugal isopotential surface, within which orbiting material is bound to a star), resulting in mass and angular momentum exchange with their companions, e.g. \cite{Paczy71}). The resultant stripped star mass donors and rejuvenated mass recipients are characterized by masses, rotational properties, and surface chemistries that are entirely at odds with evolution models of single stars \cite{Brott+2011,2021A&A...650A.107M,2018A&A...615A..78G}. 
Identifying and characterizing such systems at various stages of these processes is critical to informing models of mass and angular momentum transfer, two of the least understood processes of binary evolution.\


The rapid and violent evolution resulting from mass transfer and merger events also potentially provides fertile grounds for the generation of powerful magnetic fields \cite[e.g.][]{2015SSRv..191...77F}. Using the hot magnetic B star $\tau$~Sco as a framework, \cite{Schneider:2019} used magneto-hydrodynamic simulations and evolutionary modelling to propose that this star is a Blue Straggler, and the product of the merger of two approximately $9~M_\odot$ progenitor stars. Most significantly, they demonstrate that the magnetic field of $\tau$~Sco was plausibly generated via amplification of a seed field during the merger process. Studies by \cite{2023Sci...381..761S} and \cite{2024Sci...384..214F} presented evidence that the strongly magnetized quasi-Wolf Rayet star HD\, 45166A and the massive magnetic O-type star HD\, 148937A are also the products of recent mergers, further supporting the idea of a significant merger pathway toward magnetic field generation in massive stars.

HD 47129 \citep[Plaskett’s star;][]{1922MNRAS..82..447P} is a bright ($V = 6.06$) spatially unresolved binary system with two distinct spectroscopic O-type components. The spectroscopic component with narrower lines is in a very clear $P_{\rm orb} = 14.39626\pm 0.000953$~d orbital period radial velocity (RV) orbit \citep[e.g.][]{1922MNRAS..82..447P,Linder2008}, and the system has historically been understood to be a nearly equal-mass, non-eclipsing, very high mass double-line spectroscopic binary (SB2) system in an approximately circular orbit \citep[e.g.][]{1922MNRAS..82..447P,Stickland1987,Linder2008}. According to the analysis of \cite{Linder2008}, the system is composed of an O8III/I spectroscopic component (also referred to as ‘the narrow-line star’) of {minimum (projected)} mass of $M_1 \sin^3 i = 45.4\pm 2.4\,M_\odot$, and an O7.5V/III spectroscopic component (also referred to as ‘the broad-line star’) of minimum mass $M_2 \sin^3 i = 47.3\pm 0.3\,M_\odot$, {$i$ being the orbital inclination}. Incorporating historical estimates of the system's maximum orbital inclination led \cite{Linder2008} to absolute mass estimates of between 52 and 55~$M_\odot$ for the narrow-line component, and between 54 and 58~$M_\odot$ for the broad-line component.

In the context of the Magnetism in Massive Stars (MiMeS; \cite{2016MNRAS.456....2W}) survey, \cite{2013MNRAS.428.1686G} reported the detection of an organized magnetic field in the photosphere of the broad-line star. Following this detection it was proposed that HD\,47129 is a potential post-mass-transfer system with a magnetic field generated as a consequence \citep{2014IAUS..302....1L,Vanbeveren2018} -- but the empirical evidence supporting this was lacking, {and the then-accepted physical properties of the system were difficult to reconcile with such a scenario.}


Detailed modelling by \cite{2022MNRAS.512.1944G} established the rotational period ($P_{\rm rot, 2}=1.21551^{+0.00028}_{-0.00034}$~d) and magnetic geometry (dipole obliquity angle $\beta$ consistent with $90^\circ$ and a polar strength of about 850 G) of the magnetic broad-line star. \cite{2022MNRAS.512.1944G} also indirectly revisited the RV variability of the system. Their measurements of Zeeman Stokes $V$ circular polarization could not be reconciled with the historically-accepted large RV variability of the broad-line star, underlying a fundamental lack of understanding of the nature of Plaskett's {star and casting doubt on any previous analyses.}


Here, we analyze archival high-resolution spectra to derive the orbit and physical parameters of Plaskett's star {using a different approach to determining the stellar RVs. In contrast to previous results}, our analysis shows that the two components of the system exhibit an extreme mass ratio. This, in combination with the system's derived properties, supports the interpretation that Plaskett's star is an evolved binary system in which the now more massive, rapidly rotating magnetic component vigorously accreted mass from its narrow-line, originally more massive companion.
We further demonstrate that the system's physical, chemical, magnetic, and rotational properties can be coherently understood within that context. Specifically, we demonstrate that the magnetic field of the broad-line component likely resulted from the associated rapid mass transfer, supporting mass transfer as a pathway to magnetic-field generation in massive stars. {The emergence of magnetic fields during mass transfer substantially impacts the mass transfer process and consequently the evolution of massive binaries in general.}

\section*{Results}

\subsection*{Radial velocities and orbital analysis}
\label{sec:orb}

RVs of both components were measured from high resolution spectra obtained with the Echelle Spectropolarimetric Device for Observations of Stars (ESPaDOnS) at the Canada-France-Hawaii Telescope (CFHT; see Methods, subsection ``Observations) by computing cross-correlation functions (CCFs; \cite{Zucker1994}) of various spectral lines. {While for the narrow-line star a plethora of lines are available, for the magnetic broad-line star only the O\,{\sc iii}\,$\lambda 5592$ line appears not to be contaminated by its companion (see Methods).} The templates against which the CCFs were computed were constructed from the observations themselves by co-adding them in a common frame-of-reference \cite[e.g.][]{Shenar2019, Dsilva2020}. To obtain a first set of RVs to enable this co-addition, we use a synthetic spectrum with effective temperature $T_{\rm eff} = 32.5\,$kK and surface gravity $\log g = 4.0\,[\cms]$ for both stars, retrieved from the published grid of OB-type star spectra computed with the spectrum synthesis code TLUSTY  \cite{Lanz2003}. Using the synthetic spectrum enables us to obtain absolute RVs, which are then used to form the co-added spectrum of each component using the available spectra. We note that the physical parameters chosen for the initial TLUSTY models have a negligible impact on our measurements.

Despite the significant scatter of the RVs of the secondary, a clear RV variation in anti-phase with that of the primary is observed.

\begin{figure}
  \centering
\includegraphics[width=.5\textwidth]{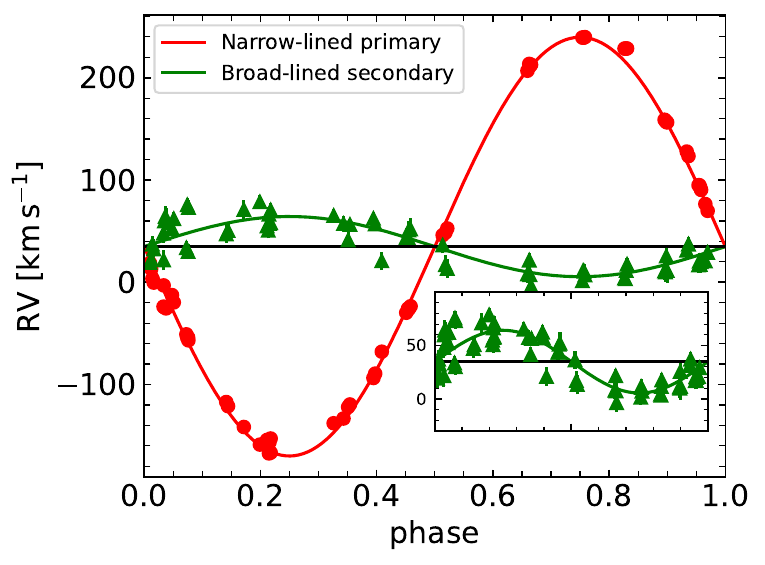}
    \caption{Orbital solution to the RVs of both components. The inset zooms on the solution of the secondary, illustrating a clear anti-phase sinusoidal variability that can be seen despite the strong variability. The RMS of both solutions are 7.6 and 11.8\,$\kms$~for the primary and secondary, respectively. {Reduced $\chi^2$ is 34.9, implying substantial RV variability of both objects beyond the formal statistical uncertainties.
    }} 
    \label{fig:OrbitSolution}
\end{figure}

We fit the orbit with the Python lmfit ({\tt https://lmfit.github.io/lmfit-py/}) package \cite{Newville2014} using the differential evolution minimization method, fitting:

\begin{equation}
    {\rm RV(\nu)}_{1, 2} = V_0 \pm K_{1, 2} \cdot \left( \cos(\omega + \nu) + e \cdot \cos \omega \right),
\end{equation}
where ${\rm RV}$ is the measured radial velocity, $\nu$ is the true anomaly, $V_0$ is the systematic velocity, $K$ is the RV semi-amplitude, $\omega$ is the argument of periastron, $e$ is the eccentricity, and the subscripts $1,2$ refer to the primary and secondary, respectively. We fix the period to $P= 14.39626\pm 0.000953\,$d as derived by \cite{Linder2008}.  Since the fitting resulted in a nearly circular orbit ($e = 0.035\pm0.005$), we fix  $e = 0$ and $\omega = 90^\circ$, such that conjunction occurs at phase $\phi = 0$ (secondary in front of the primary). The orbital elements are shown in Table\,\ref{tab:all_params}, and the orbital solution is presented in Fig.\,\ref{fig:OrbitSolution}. We also provide the projected semi-major axes $a \sin i$ and the minimum  masses $M \sin^3 i$.

The ratio of the RV semi-amplitudes implies a mass ratio $q=6.9^{+1.2}_{-0.9}$, suggesting that Plaskett's star is an extreme mass-ratio binary. 
{The orbital solution also yields  minimum  masses of $M_1 \sin^3 i = 2.4\pm 0.5\,M_\odot$ and $M_2 \sin^3 i = 16.8 \pm 0.7\,M_\odot$}. 
{
We note that all spectral lines of both components exhibit excess variability beyond the measurement noise, which may induce additional systematic errors not accounted for in the formal uncertainties. Possible causes include tidal interactions and surface deformations, stellar winds, magnetospheric effects, stellar spots \citep{Mahy2011}, or other forms of line-profile variability not captured by our RV measurement model. Consequently, the orbital solution yields a high reduced $\chi^2$ of 34.9, driven primarily by the residual scatter of the narrow-lined primary: while its formal RV measurement uncertainties are of order $1\,\kms$, the root-mean-square (RMS) of the residuals amounts to $7.6\,\kms$, indicating substantial excess scatter. Nevertheless, this scatter remains small compared to the large RV amplitude of the primary, and the derived orbital parameters are therefore well constrained. Moreover, they are consistent with the independent orbital solution of \cite{Linder2008}.
}

As projected by \cite{2022MNRAS.512.1944G}, this conclusion contrasts fundamentally with historical notions of the system \cite[e.g.][]{Linder2008}, {and strongly supports it being a post mass-transfer binary, in which the now less massive primary was originally the more massive star.  }

{The inclination of the rotation axis $i_{\rm rot, 2}$ of the rapidly-rotating secondary was previously estimated via Zeeman Doppler imaging tomographic surface mapping \cite{2022MNRAS.512.1944G} adopting various assumptions about the {(then uncertain)} RVs of the secondary. The analysis demonstrates that the resulting inclination angle is not strongly dependent on these assumptions. We adopt here the value $i_{\rm rot, 2} = 48\pm4^\circ$ obtained for constant secondary RVs, which is a good assumption given the low-amplitude RV motion measured herein. This inclination is also consistent with the value obtained from the rotational period of the star (see Results, subsection ``Physical parameters). For post mass-transfer binaries, it is expected that the rotation axis of the secondary aligns with the orbital axis {as a consequence of tidal interaction and transfer of orbital angular momentum to the rotation of the mass recipient \citep[see, e.g.][]{1980A&A....92..167H,1981A&A...102...17P,1981A&A....99..126H} leading to $i_{\rm rot, 2} = i$.} Assuming this, we perform a Monte-Carlo simulation to compute the absolute masses, semi-major axes, and Roche-lobe radii for the two components, which are given in Table\,\ref{tab:all_params}. The masses  thus derived are $M_1 = 5.9^{+1.8}_{-1.4}\,M_\odot$ and $M_2 = 40.8^{+9.3}_{-6.7}\,M_\odot$. 
The derived mass of the magnetic star is comparable to that of other known magnetic O-type stars \citep[see][]{2015ASPC..494...30W}, while the mass of the stripped primary is the highest among Galactic analogues \cite{Shenar2020_LB1, Bodensteiner+2020}.
}



\subsection*{Spectral disentangling}
\label{sec:specdis}

To investigate the RVs of the secondary using additional lines in the spectrum {beyond the O\,{\sc iii}\,$\lambda 5592$ line}, and to extract the spectra of both components from the entangled spectra, we resort to spectral disentangling.
Spectral disentangling is a method developed to separate the composite spectra of spectroscopic binaries to the constituent spectra while solving for the orbital parameters \cite{Bagnuolo1991, Hadrava1995}.

\begin{figure*}
\centering
\begin{tabular}{ccc}
\includegraphics[width=0.3\textwidth]{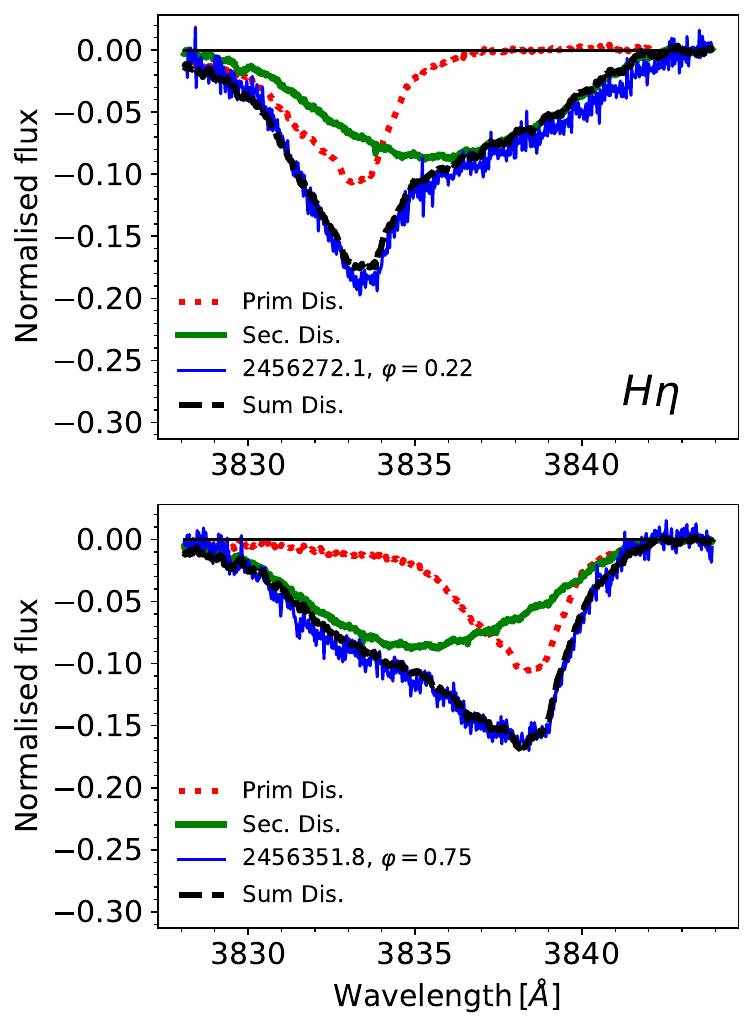} &
\includegraphics[width=0.3\textwidth]{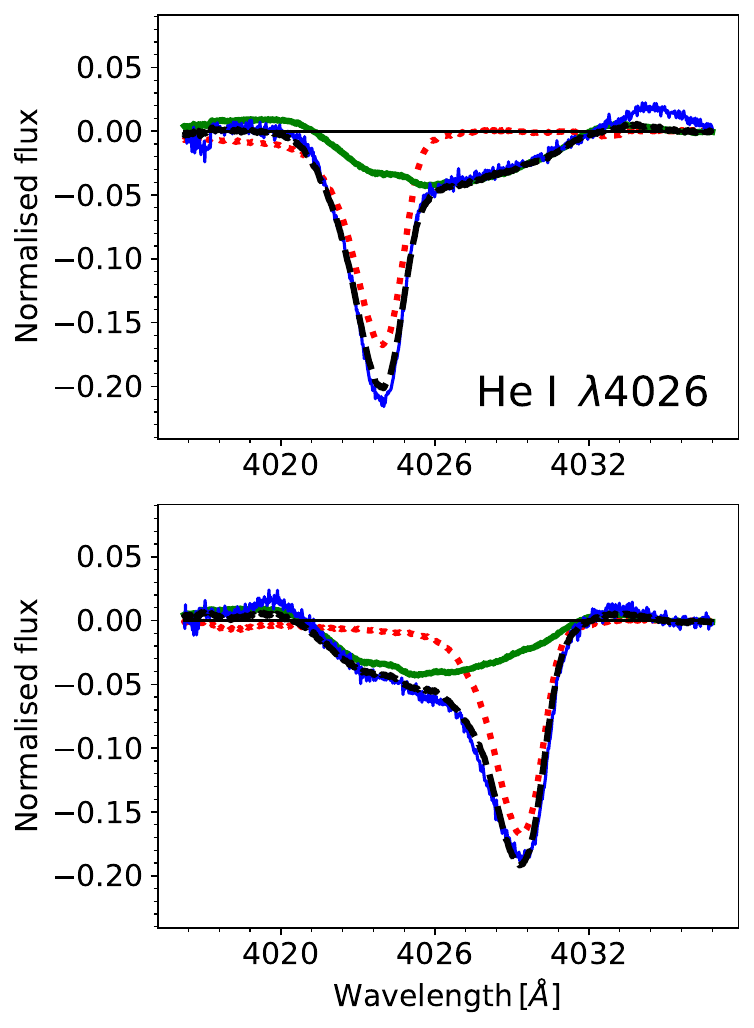} &
\includegraphics[width=0.3\textwidth]{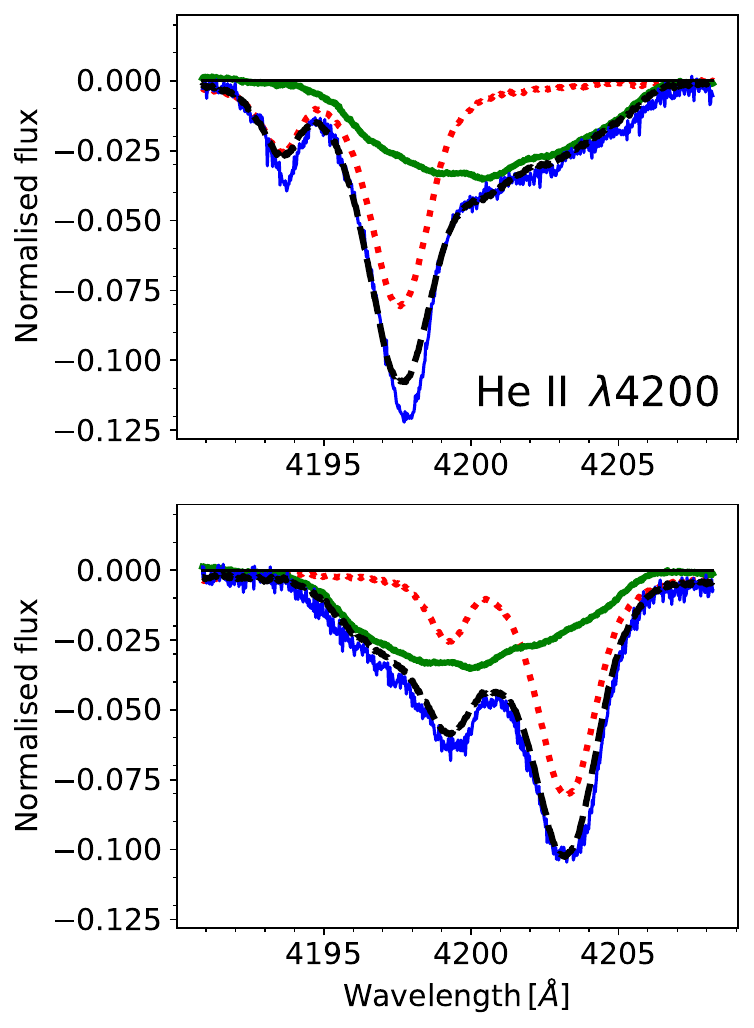}  \\
\includegraphics[width=0.294\textwidth]{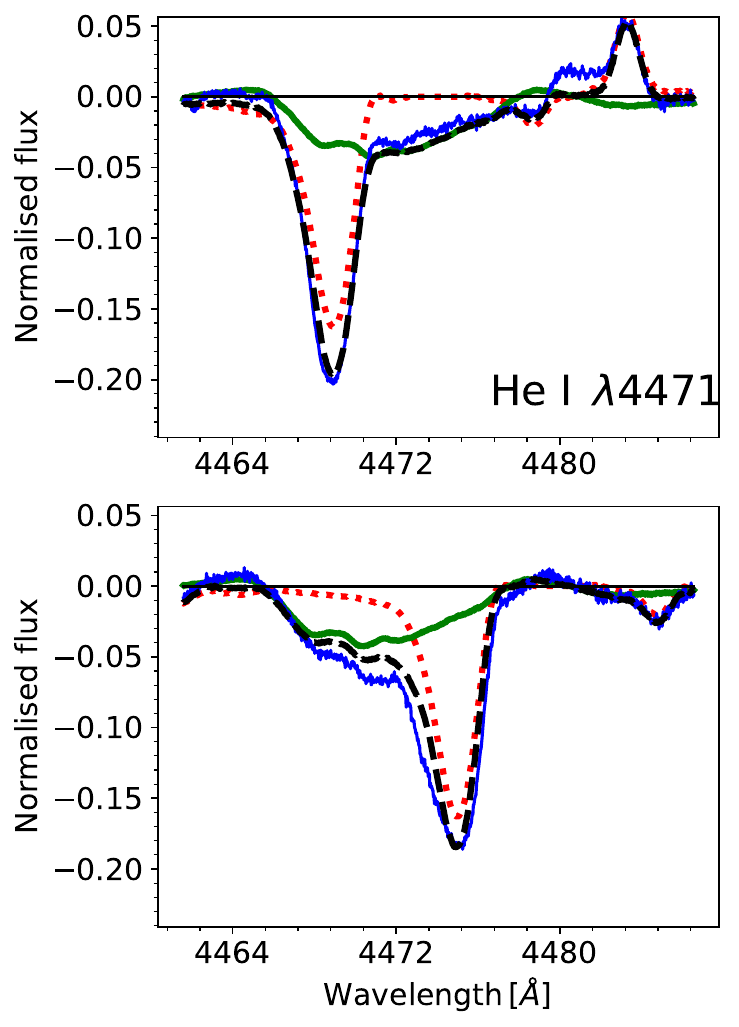} &
\includegraphics[width=0.3\textwidth]{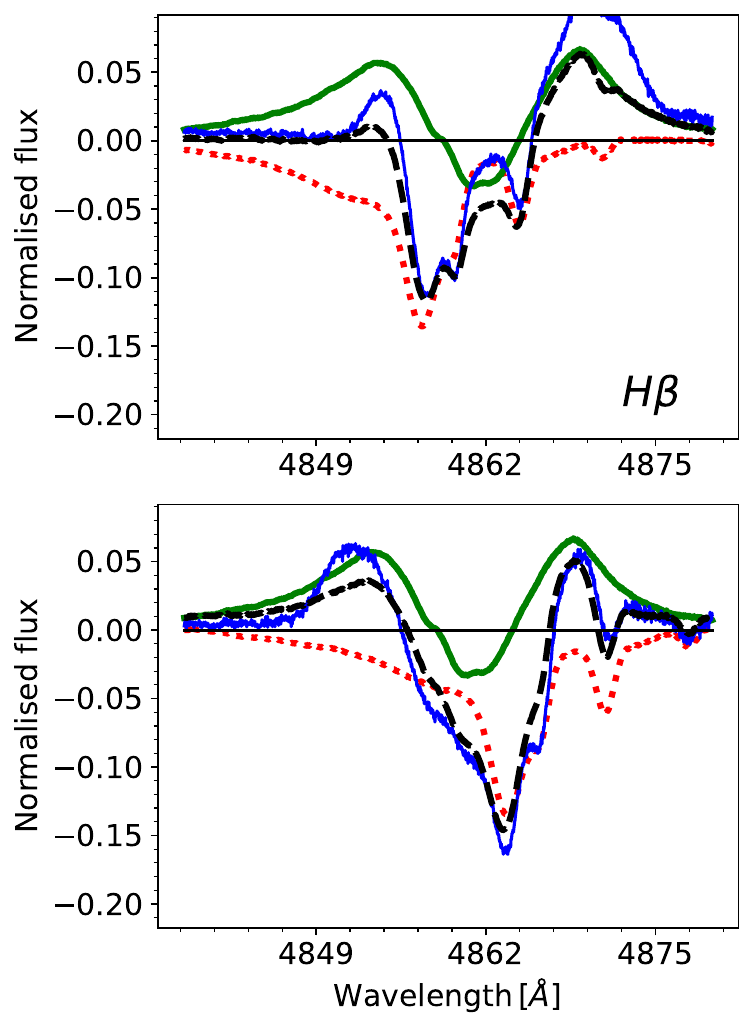} &
\includegraphics[width=0.3\textwidth]{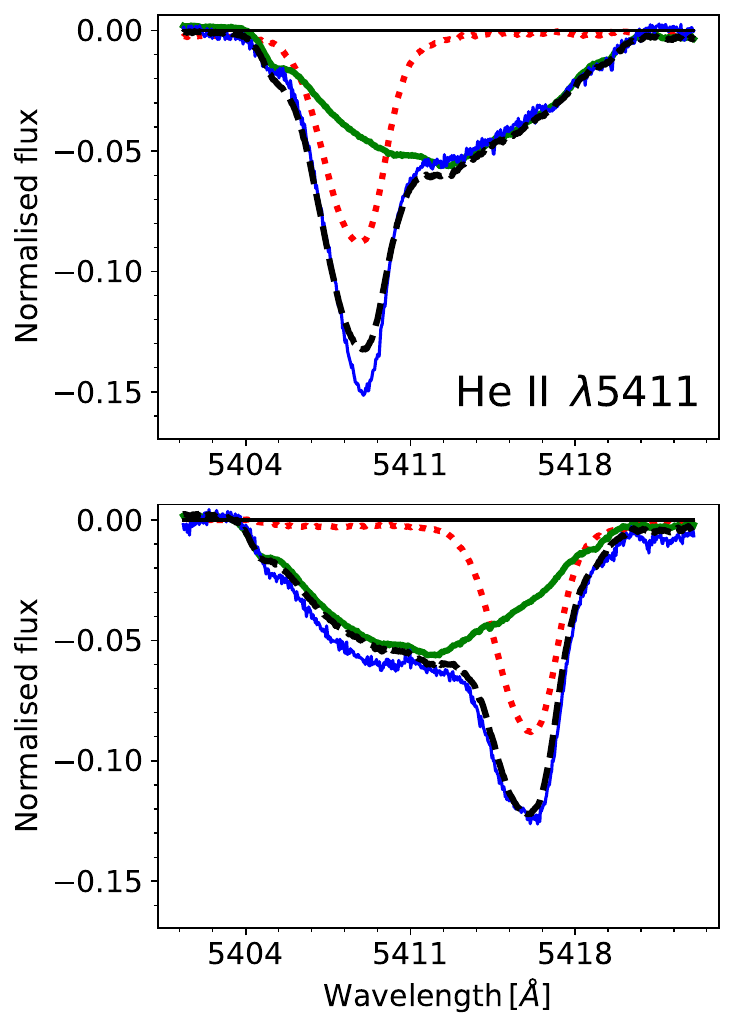}  \\
\end{tabular}
    \caption{Comparison of the observed composite spectra with the disentangled spectra of the two stellar components at two orbital phases. The narrow-lined primary is shown by a red dotted line and the magnetic broad-lined secondary by a green solid line. Their summed spectrum is shown by a black dashed line, and the observed spectrum by a blue irregular line. The six spectral regions contain H$\eta$, He\,{\sc i}\,$\lambda 4026$, He\,{\sc ii}\,$\lambda 4200$, He\,{\sc i}\,$\lambda 4471$, H$\beta$, and He\,{\sc ii}\,$\lambda 5411$. The upper and lower rows correspond to epochs near the radial-velocity extrema, at orbital phases $\phi=0.22$ and $\phi=0.75$, respectively. Line styles distinguish the components independently of colour.}
\label{fig:DisLines}    
\end{figure*}

Here, we use the shift-and-add technique \cite{Marchenko1998}, thoroughly described by \cite{Gonzalez2006, Shenar2020_LB1, Shenar2022} -- see Methods.
This technique was already implemented on Plaskett's star by \cite{Linder2008} to compute the disentangled spectra and the resulting RVs for the secondary. However, the disentangling implemented by \cite{Linder2008} relied on an initial RV measurement using the CCFs of blended lines. We argue that these RV measurements were biased since they were performed on { lines  that are strongly contaminated by (variable) magnetospheric emission and that remain blended during the entire orbital cycle (see also Sect.\,\ref{sec:mismatch_orbit}).}  This introduces considerable degeneracy between the shape, variability, and centre-of-gravity motion of the magnetic star's spectral lines.  When disentangling assumes similar initial RV amplitudes for both components, this degeneracy propagates into the disentangling, which in turn propagates to the RV measurement of the next iteration.  Our numerical experiments lead us to believe that this is the origin of the large value of $K_2 = 192.4\pm6.7\,$\kms~derived by \cite{Linder2008}.

Here, we do not adopt the RVs of the secondary {a priori}, but derive $K_2$ independently via disentangling.
The results of the new disentangling (shown in Fig.\,\ref{fig:DisLines}) are in good agreement with the orbital RV analysis presented in Sect.~\ref{sec:orb}. The lines H$\eta$ ($K_2 = 9.8\pm7.1\,$\kms), He\,{\sc i}\,$\lambda 4026$ ($K_2 = 15.5\pm10.7\,$\kms), He\,{\sc ii}\,$\lambda 4200$ ($K_2 = 26.0\pm9.9\,$\kms), He\,{\sc i}\,$\lambda 4472$ ($K_2 = 25.5\pm10.7\,$\kms), and He\,{\sc ii}\,$\lambda 5411$  ($K_2 = 10.6\pm7.6\,$\kms) agree within $1-2\sigma$ with our $K_2$ measurement obtained from the orbital solution ($K_2 = 29.6\pm4.5\,$\kms). The mass ratio obtained from the weighted mean of all helium lines (omitting Balmer lines due to contamination with emission) is $q=11.4^{+4.0}_{-2.4}$, consistent with the value of $q=6.9^{+1.2}_{-0.9}$ within 2$\sigma$.
Generally, the results are found to depend sensitively on the exact disentangled region and normalization. {It is furthermore possible that our quoted uncertainties are underestimated due to systematics originating from stochastic variability (see Sect.\,\ref{subsec:specdissupp})}. We therefore deem the results obtained via direct RV derivation via CCF (Sect.\,\ref{sec:orb}) 
more robust, and only provide $q$ from disentangling  to demonstrate that the interpretation of the system as an extreme mass-ratio binary is valid regardless of the line and method used. Importantly, the value of $q = 1.04 \pm 0.05$ reported by \cite{Linder2008} is fully rejected ($> 6\sigma$ discrepancy) by our analysis, in agreement with \cite{2022MNRAS.512.1944G}. 

{ Both components show significant emission in the disentangled spectra. While this emission could be affected by spectral variability and the limitations of the method, the fact that both components exhibit clear emission features associated with them in the raw data gives credibility to the results. The broad-line secondary shows double peaked emission in hydrogen Balmer lines and He\,{\sc ii} lines, likely originating from its magnetosphere or a circumstellar disk. The narrow-line primary shows emission in a few metal lines (e.g., some N\,{\sc ii} lines). It also shows highly asymmetric, P-Cygni like line profiles in He\,{\sc ii}\,$\lambda 4686$ and He\,{\sc i}\,$\lambda 5876$, in addition to signs for emission infilling in the cores of Balmer lines, as implied from their weakness compared to our models (Sect.\,\ref{sec:spectroscopy}). In addition, available UV data (see Methods section) show a few P-Cygni lines which appear to move with the narrow-lined primary. These are tell-tale signs of a stellar wind belonging to the stripped primary.}

\subsection*{Physical parameters and abundances}
\label{sec:spectroscopy}

We perform a quantitative spectral analysis of both components using the Potsdam Wolf-Rayet (PoWR) model atmosphere code \cite{Graefener2002, Hamann2003, Sander2015, PoWRZenodo}, which solves the radiative transfer problem in radially expanding atmospheres while relaxing the assumption of local thermodynamic equilibrium (non-LTE). The object was analyzed spectroscopically in the past by \cite{Linder2008}, who misinterpreted the two components to be of similar mass given their similar spectral types, which biased the analysis. We therefore repeat the analysis here, but use their derived parameters as a starting guess for our models, with a very different outcome.

\begin{table}[!t]
\centering
\caption{Orbital and physical parameters derived or adopted for the system. Symmetric errors are $1\sigma$ (68\%), non-symmetric errors are 68\% credible intervals (16$^{\rm th}$ and 84$^{\rm th}$ percentiles) given around medians of posteriors.}
\label{tab:all_params}
\renewcommand{\arraystretch}{1.2}
\setlength{\tabcolsep}{4pt}
\begin{tabular}{lcc}
\hline\hline
 & Narrow-line primary & Broad-line secondary \\
\hline
\multicolumn{3}{c}{\textit{Orbital elements and mass constraints}} \\
\hline
$P$ [d]\tnote{a}         & \multicolumn{2}{c}{$14.39626 \pm 0.000953$} \\
$T_0$ [MJD]              & \multicolumn{2}{c}{$55966.19 \pm 0.06$} \\
$V_0$ [$\kms$]           & \multicolumn{2}{c}{$34.8 \pm 1.1$} \\
$K$ [$\kms$]             & $204.7 \pm 1.6$ & $29.6 \pm 4.5$ \\
$\omega$ [$^\circ$]       & \multicolumn{2}{c}{90 (fixed)\tnote{b}} \\
$e$                       & \multicolumn{2}{c}{0 (fixed)\tnote{b}} \\
$q=M_2/M_1$ (RVs)        & \multicolumn{2}{c}{$6.9^{+1.2}_{-0.9}$} \\
$q$ (disentangling)      & \multicolumn{2}{c}{$11.4^{+4.0}_{-2.4}$} \\
$M \sin^3 i$ [$M_\odot$] & $2.4 \pm 0.5$ & $16.8 \pm 0.7$ \\
$a \sin i$ [$R_\odot$]   & $58.4 \pm 0.5$ & $8.4 \pm 1.3$ \\
$i$ [$^\circ$]\tnote{c}  & \multicolumn{2}{c}{$48 \pm 4$} \\
$M$ [$M_\odot$]          & $5.9^{+1.8}_{-1.4} $ & $40.8^{+9.3}_{-6.7}$ \\
$a$ [$R_\odot$]          & $78.4^{+5.5}_{-4.5}$ & $11.4^{+1.9}_{-1.8}$ \\
$R_{\rm RL}$ [$R_\odot$] & $20.6^{+1.9}_{-1.7}$ & $49.3^{+3.5}_{-2.8}$ \\
\hline
\multicolumn{3}{c}{\textit{Physical parameters and abundances}} \\
\hline
Distance [pc]\tnote{d}            & \multicolumn{2}{c}{$1270 \pm 96$} \\
$E_{B-V}$ [mag]          & \multicolumn{2}{c}{$0.395 \pm 0.006$} \\
$R_{V}$ [mag]          & \multicolumn{2}{c}{$3.29 \pm 0.12$} \\
$l(V)$                   & $0.550 \pm 0.050$ & $0.450 \pm 0.050$ \\
$T_{\rm eff}$ [kK]        & $31.0 \pm 1.0$ & $34.0 \pm 2.0$ \\
$\log g$ [\cms]          & $3.3 \pm 0.2$ & $4.0$ (fixed) \\
$\log L$ [$L_\odot$]     & $5.10 \pm 0.07$ & $5.12 \pm 0.11$ \\
$R$ [$R_\odot$]          & $12.3 \pm 1.2$ & $10.4 \pm 1.4$ \\
$\log \dot{M}$ [$M_\odot\,{\rm yr}^{-1}$]          & $-5.8 \pm 0.3$  & - \\
$v_\infty$ [\kms]          & $2300 \pm 100$  & - \\
$R/R_{\rm RL}$           & $0.59 \pm 0.08$ & $0.21 \pm 0.03$ \\
$v \sin i$~[\kms]           & $53.2 \pm 2.7$ & $350 \pm 50$ \\
$X_{\rm H}$              & $0.40 \pm 0.10$ & solar (0.73) \\
$X_{\rm C}/10^{-3}$      & $0.090 \pm 0.045$ & solar (2.3) \\
$X_{\rm N}/10^{-4}$      & $34 \pm 17$ & solar (6.93) \\
$X_{\rm O}/10^{-3}$      & $< 0.05$ & solar (5.77) \\
\hline
\end{tabular}
\begin{tablenotes}
\item[a] Period is fixed according to \cite{Linder2008}.
\item[b] Allowing for an eccentric solution resulted in $e = 0.035\pm0.005$ and $\omega = 22\pm8^\circ$.
\item[c] Assuming the orbital inclination is aligned with the rotation axis of the secondary derived by \cite{2022MNRAS.512.1944G}.
\item[d] Adopted from \cite{Bailer-Jones2021}, but see text for potential caveats.
\end{tablenotes}
\end{table}

For the analysis, we use the disentangled spectra obtained in Sect.\,\ref{sec:specdis}. The spectra need to be scaled in opposite proportion to the light contribution of each component in the visual, $l(V)$, to account for line dilution.  We estimate the light contribution of the narrow-line primary by inspecting several metal lines belonging to S, Si, and Mg, whose abundances are assumed to be solar. {The latter assumption is motivated by the fact that these elements are not expected to be significantly affected by the evolutionary processes relevant for the primary, unlike, for example, CNO elements.} This analysis yields an estimated light contribution of $l(V)_1 = 55\pm 5\% $, implying $l(V)_2 = 45\mp 5\% $ for the much more massive secondary.  {The uncertainty reflects the range of light ratios that provide an acceptable visual match to the observed metal-line strengths. The light ratio impacts primarily the derived luminosities, but those  do not affect our conclusions regarding the evolutionary status of the system, which are primarily based on the orbital analysis.}

The effective temperature and surface gravity of the primary are estimated from the ionization balance of metal lines and the strengths and profiles of the Balmer lines (see Methods section), respectively, and are estimated at $T_{\rm eff, 1} = 31.0 \pm 1.0$\,kK and $\log g_1 = 3.3\pm0.2\,$[\cms]. { We cannot reproduce the observed emission line profiles with standard wind models. However, reproducing the line widths and overall strengths implies a terminal wind velocity of $v_\infty = 2300\pm100\,$\kms~and a mass-loss rate of $\log \dot{M} = -5.8\pm0.3$\,$[M_\odot\,{\rm yr}^{-1}]$. 
}

For the broad-line secondary, all Balmer lines are strongly contaminated with emission, preventing the derivation of $\log g$. We therefore assume a standard value of $\log g_2 = 4.0 \,$[\cms]  for { dwarfs} \citep{Martins2005}, which is also consistent with its derived mass (Sect.\,\ref{sec:orb}) and radius (derived below). { While deviations could be expected in the case of a potential mass accretor, the  value of $
\log g_2$ does not impact our conclusions significantly.} The effective temperature of the secondary is determined mainly from the ionization balance of He\,{\sc i}/{\sc ii}, yielding $T_{\rm eff,2} = 34.0 \pm 2.0\,$kK.  The latter value may be underestimated by up to about 2\,kK due to the rapid rotation of the star (see Methods section). 

The projected rotation velocity $v \sin i$ is estimated from the profile shape of isolated spectral lines using a Fourier analysis and goodness-of-fit method \cite{Simon-Diaz2007, Simon-Diaz2014, IACOBURL}, and is found to be $v \sin i_1 = 53.2 \pm 2.7$\,\kms~and $v \sin i_2 = 350 \pm 50$\,\kms~for the primary and secondary, respectively.

{ No significant deviations from a solar abundance pattern are observed for the broad-line secondary, though an estimation of its CNO content is hindered by strong contamination from the magnetosphere and smearing of its metal lines due to rapid rotation.} The narrow-line primary shows clear evidence for CNO-processed material and helium enrichment in its outer layers (see Methods section). The He mass fraction is estimated to be $X_{\rm He} = 0.6\pm0.1$, that is, a factor two increase in mass fraction compared to the solar baseline. The nitrogen abundance is estimated to be roughly five times the baseline value, while carbon is at about 3\% solar. For oxygen, only an upper limit of 1\% solar can be derived, consistent with the lack of detectable contribution to the O\,{\sc iii}\,$\lambda 5592$ line. The presence of CNO-processed material on the surface of the narrow-line primary is consistent with the picture of it being an evolved, post mass-transfer mass donor, as outlined in Results, subsection ``Evolutionary model.


\begin{figure}
  \centering
\hspace{-0.25cm}
\includegraphics[width=9cm]{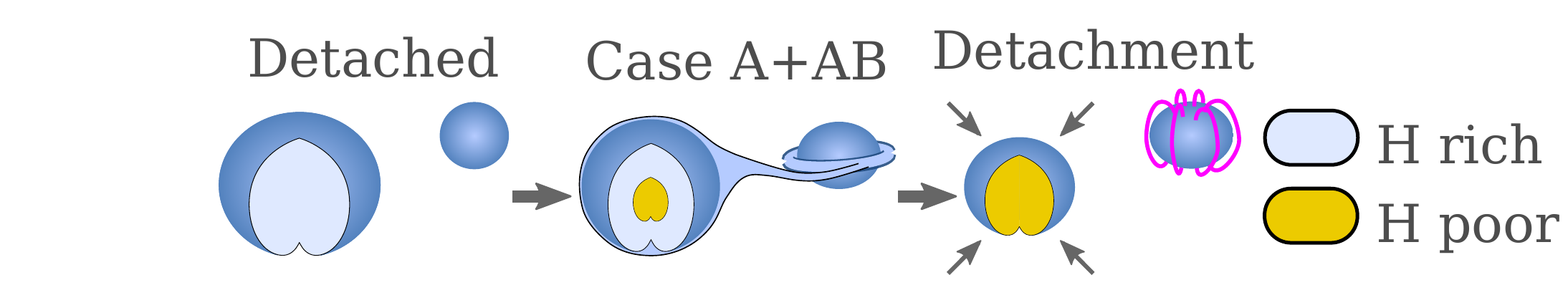}
\includegraphics[width=9cm]{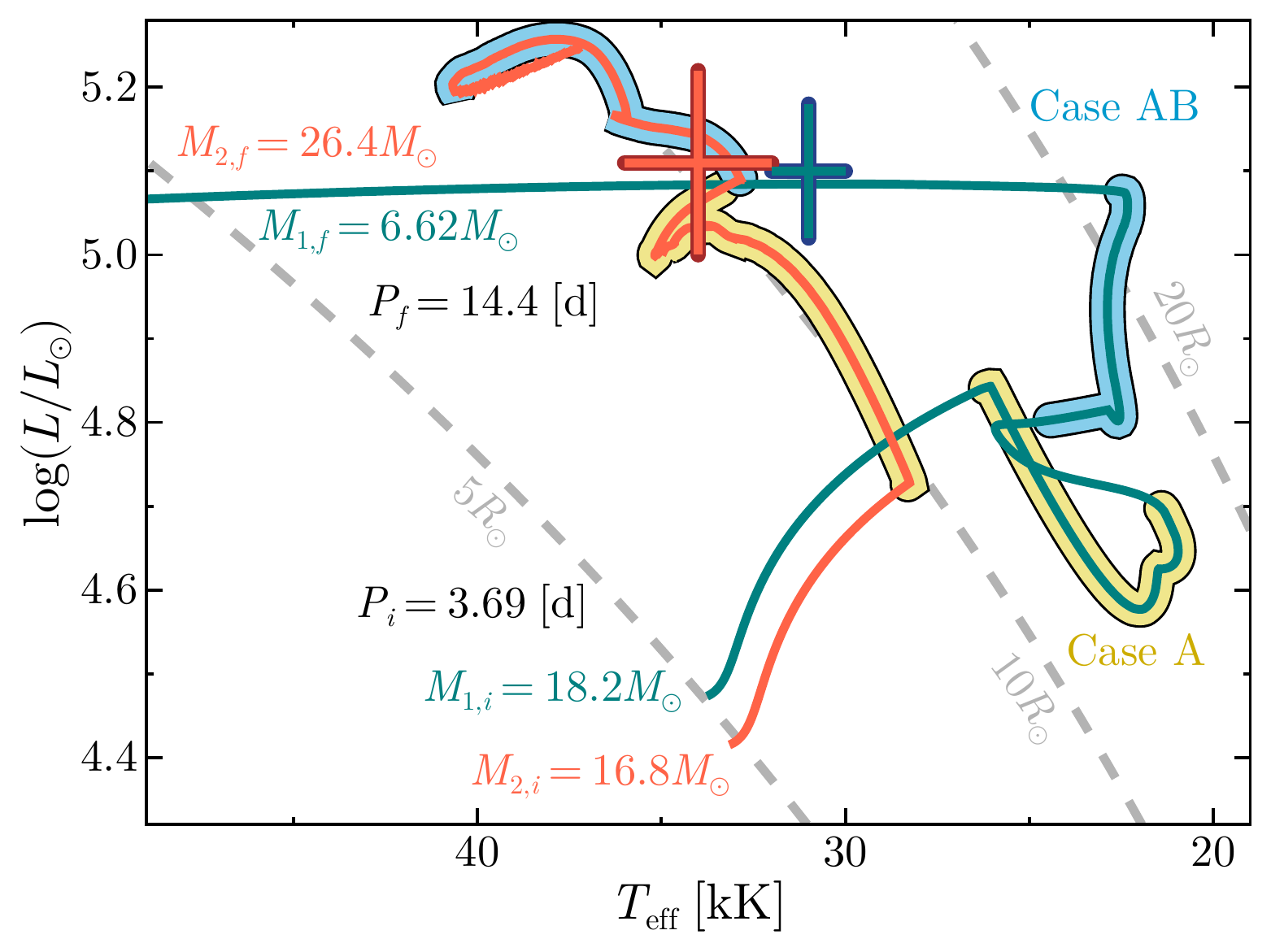}
    \caption{Illustration of the binary evolutionary model. The upper schematic identifies the successive evolutionary stages: detached evolution, mass transfer during the donor's main-sequence phase (case A), subsequent detachment, and further stripping after the main-sequence phase (case AB). The lower panel shows the corresponding evolution in the Hertzsprung--Russell diagram, with the evolutionary stages distinguished by line colour. Teal and orange crosses mark the observed effective temperatures and luminosities of the narrow-lined primary and broad-lined secondary, respectively. The model illustrated in the figure begins with primary and secondary masses of $18.2~M_\odot$ and $16.8~M_\odot$, respectively.}
    \label{fig:HR_evo}
\end{figure}

We fit the spectral models to the observed spectral energy distribution (SED) to derive the color excess $E_{B-V}$, total-to-selective extinction ratio $R_V$, and luminosities, for an adopted distance of $d = 1270 \pm 96\,$pc \citep{Bailer-Jones2021}. { Distance uncertainties  might be strongly underestimated due to the brightness of the source and its location in the Galactic plane \cite{2025MNRAS.543...63P}, though our adopted value is consistent within 1$\sigma$ with the distance  to the Mon OB2 association ($1.55\pm0.15\,$kpc, \cite{Martins2012}) in which HD~47129 is thought to reside.}
The luminosities and temperatures result in the stellar radii $R$ for both components, which in turn also enable the calculation of the spectroscopic masses $M_{\rm spec} = G^{-1}\,g\,R^2$  {($G$ is the gravitational constant)} for the primary (the secondary's $g$ could not be derived), albeit with large uncertainties. The spectroscopic mass of the primary ($M_{\rm spec, 1} = 11 \pm 5~M_\odot$) is over a factor two smaller compared to what is expected \citep{Martins2005} for an O8-type star ($27\,M_\odot$), and is consistent within $1\sigma$ with the value of $M_1 = 5.9^{+1.8}_{-1.4}\,M\odot$ derived from the orbital analysis. 
The radii further imply that both components are currently well within their Roche lobes, with the Roche-lobe filling factors $R/R_{\rm RL} = 0.59\pm0.08$ and $0.21\pm0.03$ for the primary and secondary, respectively. { While the luminosity and implied spectroscopic mass and mass-loss rate rely on our assumed distance, our conclusion for the nature of this system is robust against distance uncertainties, as are the masses derived via orbital analysis.}

The published rotational period of the broad-line component ($P_{\rm rot, 2}=1.21551$\,d \cite{2022MNRAS.512.1944G}), combined with its radius derived here ($10.4\pm1.4\,R_\odot$), implies that its equatorial rotation velocity is $v_{\rm eq, 2} = 430\pm60\,$\kms. Comparing this to the projected rotation velocity of $v \sin i_2 = 350\pm50$\,\kms~measured here yields  an inclination of the rotation axis of $i_{\rm rot, 2}=54^{+29}_{-13}\,^\circ$. This value is in agreement with $i_{\rm rot, 2}=48\pm4^\circ$  derived by \cite{2022MNRAS.512.1944G}, which we adopted for the absolute mass calculation in Results, subsection ``Radial velocities and orbital analysis.

\subsection*{Evolutionary model}
\label{sec:ev}

In order to produce a model that can explain the essential properties of Plaskett's star, we make use of the binary capabilities of the MESA stellar evolution code \citep{Paxton+2011,Paxton+2015} using the physical assumptions for wind mass loss and composition given for a Galactic environment by \cite{Brott+2011}. Owing to the large present-day mass ratio we assume fully conservative mass transfer and compute our models without rotation, as there are significant uncertainties concerning the efficiency of mass transfer and accretion spin-up { (see \cite{Lechien2025, Picco2026} and} the discussion by \cite{MarchantBodensteiner2024}). {Full details of our simulation setup are provided in Methods, subsection ``MESA simulation setup}. Possible solutions could be found by exploring the full set of initial parameters (assuming an initially circular orbit) which are the initial mass of the primary $M_{1,\mathrm{i}}$ (which becomes the narrow-line non-magnetic stripped star), the initial mass ratio $q_\mathrm{i}\equiv M_{2,\mathrm{i}}/M_{1,\mathrm{i}}$ and the initial orbital period $P_\mathrm{i}$. Instead, as we assume fully conservative mass transfer, we make use of the analytical relationships that relate the orbital period $P$ and mass ratio $q$ to their initial values during a mass transfer phase (e.g. \cite{Soberman+1997}),
\begin{equation}
    \frac{P}{P_\mathrm{i}}=\left(\frac{q}{q_\mathrm{i}}\right)^{-3}\left(\frac{1+q}{1+q_\mathrm{i}}\right)^6. \label{equ:orbevo}
\end{equation}
Considering the observed present-day mass ratio $q$ of nearly 7 and period $P= 14.4\;\mathrm{d}$, this expression relates the initial mass ratio and period, reducing the dimensionality of the space that needs to be explored (with only a small error due to ignoring wind mass loss in Equation \ref{equ:orbevo}). Choosing initial conditions based on Equation \ref{equ:orbevo} does not ensure that the simulation will reach the desired mass ratio and period, {as the donor can complete its stripping and detach before that. Only if stripping proceeds until matching the target mass ratio will the orbital period match the observed one.}

{Using simulations with initial total masses comparable to the present day and different initial mass ratios, no solutions were found that match a mass ratio of nearly 7 as determined through the RVs. Owing to this we computed a grid of simulations where we varied the total initial mass, the initial mass ratio $q_i$ and also considered lowered present day mass ratios $q$ in Equation (\ref{equ:orbevo}). One example simulation from our grid that matches the present day luminosities but has a lower total mass and mass ratio than determined from the observations is shown in Figure \ref{fig:HR_evo}. After stripping the donor becomes a $6.62~M_\odot$ helium rich star while the accretor grows to $26.4~M_\odot$ (corresponding to $q=4$) and an orbital period of $14.4$ days. {The accretor also becomes enriched with products of CNO burning, becoming helium ($X_\text{He}=0.42$) and nitrogen rich ($X_\text{N}/10^{-4}=27$)}. Both stars' luminosities are reproduced within their $1\sigma$ errors, but the final temperature of the accretor is higher than observed, at $40$~kK. This can be partly resolved by accounting for rotation: artificially adding uniform rotation reaching 90\% of the critical rate to this accretor lowers its effective temperature by $2~\mathrm{kK}$, while accounting for rotation in the atmosphere analysis can cause an increase in the derived effective temperature. }

{Even though our model can reproduce the observed luminosities and orbital period, the extreme mass ratio of Plaskett's star represents a significant challenge to evolutionary models, analogous to the outcome of attempts to reproduce the properties of the recently stripped star system HR~6819 \cite{Bodensteiner+2020}}









\section*{Discussion}

\subsection*{Summary}

{Our model of the Plaskett's Star system requires a hot (31~kK), intermediate-mass ($5.9^{+1.8}_{-1.4}~M_\odot$) companion to the magnetic broad-line star (34~kK, $40.8^{+9.3}_{-6.7}~M_\odot$) to explain the large ratio of RV amplitudes. The non-magnetic companion exhibits extreme light-element chemical peculiarities, including enhanced helium and nitrogen, and deficient carbon and oxygen (while the chemistry of the magnetic star appears substantially solar). While the rotation of the companion is {modest}, the magnetic star has a projected rotational velocity {$v_{\rm eq, 2}$ of approximately 430~km\,s$^{-1}$}, approximately 60\% of critical, corresponding to a rotational period of 1.21 days. We interpret the companion to be a stripped star, and the remnant of a massive companion that once dominated the mass budget of the binary. As that companion evolved off the main sequence, Roche Lobe overflow (RLOF) occurred, leading to mass transfer onto the broad-line star. Modelling suggests that for nominal initial masses of the two components of 18.2 and 16.8~$M_\odot$ in a 3.69-d orbit yield present-day masses of approximately 6.6~$M_\odot$ for the stripped primary and 26~$M_\odot$ for the secondary, with an orbital period compatible with that observed. Our model is unable to reproduce the extreme ($q$ of nearly 7) mass ratio implied by the observations while simultaneously reproducing the orbital and physical properties of the system.} {This matter is discussed further in Methods, subsection ``On the discrepancy between the observed mass ratio and that predicted by evolutionary modeling.}

{ 

\subsection*{Formation and properties of the stripped star}

Plaskett's star joins the sparse sample of systems similar to LB-1 \cite{Shenar2020_LB1} and HR~6819 \cite{Bodensteiner+2020}, in which a "bloated stripped star" is observed during the rapid contraction towards the helium main sequence following a rapid mass-transfer phase. Its mass, however, is the largest observed for its type in the Galaxy, significantly greater than the Chandrasekhar mass limit (1.4\,$M_\odot$) required for it to undergo core-collapse in a few Myr. Its high mass is rivaled only by analogous systems found in the Small and Large Magellanic Clouds \cite{Ramachandran2023, Villasenor2023}. The fact that the stripped star exhibits substantial hydrogen in its envelope is in agreement with other observations of bloated stripped stars, but at odds with hot stripped stars recently identified in the Magellanic Clouds \cite{Drout2023, Goetberg2023}, which tend to be hydrogen free. 

Interpreting the emission features of the  bloated stripped star as wind features, the spectral analysis also enables us to estimate the wind mass loss of stars in this observationally rare, yet evolutionarily common phase of massive stars. While subject to uncertainties due to possible distance error underestimation and contamination from the companion, our model suggests a mass-loss rate of $\log \dot{M} =-5.8\pm0.3$\,$[M_\odot\,{\rm yr}^{-1}]$. The fact that the star exhibits a significant wind could be anticipated from its relative proximity to the Eddington limit. The luminosity, dynamical mass, and hydrogen mass fraction imply an Eddington factor (i.e., the ratio of radiative acceleration due to electron scattering to gravitational acceleration) of $\Gamma_e = 0.46$.

To our knowledge, there are no predictions for mass-loss rates of bloated stripped stars. {The derived mass-loss rate lies well above predicted mass-loss rates for main-sequence stars of this spectral type of approximately $\log \dot{M}=-6.6$\,$[M_\odot\,{\rm yr}^{-1}]$ \cite{Vink2001}, but below Wolf-Rayet mass-loss rates in excess of $\log \dot{M} =-5.0$\,$[M_\odot\,{\rm yr}^{-1}]$ \cite{Hamann1995}, as could be expected for an intermediate-mass stripped star.} Theoretical predictions for mass-loss rates of stripped stars for a fixed temperature of 50\,kK were computed by \cite{Vink2017}. For the measured (distance-dependent) luminosity of the stripped primary in Plaskett's star, this prescription yields $\log \dot{M}_{\rm pred} = -6.4\,[M_\odot\,{\rm yr}^{-1}]$, which is a factor of four smaller than observed. However, given the temperature mismatch, it is difficult to draw firm conclusions regarding the validity of the models. Regardless, the measured mass-loss rate of the stripped component suggests that winds play a significant role in shaping the final envelope properties of the stripped star, and that it may lose its entire hydrogen envelope, in agreement with the hydrogen-free atmospheres of compact hot stars recently found in the Magellanic Clouds \cite{Drout2023, Goetberg2023}.

}





\subsection*{Rapid rotation of the magnetic star}

The rotation of the Plaskett's Star secondary is entirely at odds with the rotational properties of other known magnetic O-type stars (e.g. \cite{2015ASPC..494...30W, 2017MNRAS.465.2432G}), none of which have established rotational periods shorter than 7~d,  the majority have periods longer than {one month, and three} have periods between 1.5 and 55 years. {This remarkable characteristic is generally understood as a natural consequence of magnetic braking: the efficient shedding of rotational angular momentum via the coupling between the star's magnetic field and its dense stellar wind \citep[e.g.][]{1967ApJ...148..217W,2009MNRAS.392.1022U}. The binary mass-transfer scenario proposed here} provides a natural explanation for the current anomalous rapid rotation of the secondary \citep{Cantiello:2007}, which we estimate to be $v/v_{\rm crit}$ of about 60\% critical). 

The fact that the magnetic star is rotating at less than critical velocity implies that some spin-down has likely occurred since mass transfer ceased. This spin-down could result from internal restructuring of the star (i.e. changes in moment of inertia) or angular momentum loss due to unmagnetized mass loss, and/or magnetic braking. The magnetic braking spin-down timescale of Plaskett's magnetic star (as computed according to \cite{2009MNRAS.392.1022U}, and approximating the mass-loss properties of the magnetic star assuming the prescription of \cite{Vink2001} is roughly 13~Myr. This rather long spin-down timescale { - due to secular braking via the wind - is challenging to reconcile with any significant magnetic braking since mass transfer ceased. Another possibility is that there was a stronger (equatorial) wind in the early post-mass-transfer stage that could have resulted in a short-term episode of accelerated spin-down.}

\subsection*{Origin of the magnetic field, and implications on stellar evolution}

Strong magnetic fields at the surfaces of massive stars are rare \citep{2015ASPC..494...30W}; they are even rarer in massive stars situated in close binaries \citep{2015IAUS..307..330A}. Given the relatively unlikely presence of the magnetic fields in the Plaskett's Star broad-line component, we are motivated to explore if its origin has any connection to the inferred mass transfer. {In this scenario we assume that the broad-line component was initially unmagnetized, and acquired its magnetic field as a consequence of binary evolution.}

There are two options for the origin of the field in the accretor: either the field was present before mass transfer, or it was generated during the accretion process. A fraction of about 10\% of massive stars are known to host large-scale magnetic fields with strengths ranging from a few hundred G to several kG \cite{2016MNRAS.456....2W}. These ``fossil", or remnant fields (i.e. fields not currently generated by a dynamo) are expected to be arranged in stable configurations consisting of interlocked poloidal and toroidal fields \cite{Braithwaite2004}, superficially consistent with the magnetic field of the broad-line star. 
{Following the literature on accretion onto magnetized
neutron stars \cite{Wang:2016}, we first estimate the field strength
required for the magnetosphere to withstand the ram pressure of the
accretion flow at the stellar surface, i.e. for the magnetospheric
radius to exceed the stellar radius. We assume a steady accretion flow
with a constant mass transfer rate $\dot{M}$, and we neglect the
effects of thermal pressure, magnetic diffusivity, and magnetic
reconnection. The cross-sectional area of the flow can be expressed as
$A_{\text{impact}} = \alpha R^2$, where $R$ is the radius of the
accretor and $\alpha < 1$ depends on the flow geometry. Under these
simplifying assumptions, in Methods, subsection ``Calculation of the critical field we derive the
critical magnetic field $B_c$ required to channel the accretion flow,
\begin{equation}\label{eq:critical}
B_c = \sqrt{\frac{8\pi \dot{M} \sqrt{2GM}}{\alpha R^{5/2}}}
\end{equation}
where $M$ is the mass of the accretor and $G$ is the gravitational constant. For our model of Plaskett's
star, when the secondary starts accreting it has $M$ of approximately
$16\,M_\odot$ and $R$ of $9\,R_\odot$. The accretion rate is approximately
$\dot{M}=10^{-4}\,M_\odot/\text{yr}$, resulting in a critical
magnetic field $B_c$ between 6 and 60~kG (for $\alpha = 1$--$0.01$).
Fields weaker than $B_c$ -- which, since we expect the flow to be
collimated ($\alpha < 1$), include all but the very strongest fields
observed in magnetic OB stars -- are directly overwhelmed by the flow
at the surface. Crucially, however, even fields exceeding $B_c$ are
not expected to avoid burial. As established for accreting neutron
stars, magnetically channeled material accumulates at the polar caps
and spreads equatorward, dragging the field lines with it and
screening the field once the accreted mass exceeds a small critical
value \citep[e.g.][]{PayneMelatos2004, Wang:2016} -- negligible
compared to the roughly $10\,M_\odot$ accreted by the secondary in our
model. Indeed, resisting burial beneath such an envelope would require
field strengths of order $10^7$~G
(Sect.~\ref{bc_calculation}), orders of magnitude beyond any field
observed in a non-degenerate star. Burial is therefore unavoidable
regardless of the initial field strength. And once buried underneath
a few solar masses of stellar material, the timescale for the field
to diffuse back out is set by Ohmic diffusion, which is far longer
than the lifetime of a massive star \citep[see, e.g. equation 5
of][]{2011A&A...534A.140C}. In accreting massive stars, buried
magnetic fields stay buried.}


This makes the scenario in which the field observed in the secondary is a fossil field that survived accretion very unlikely. 

Given the star that recently experienced accretion is magnetized, it is natural to discuss the possibility of an accretion-induced dynamo. {Mass transfer provides a natural source of free energy for magnetic-field amplification because the accreted material carries substantial angular momentum. Whether the flow forms an accretion disk or impacts the star directly, angular momentum is deposited in the outer layers of the accretor and can drive differential rotation and shear. Such differentially rotating, conducting flows are natural sites for magnetic-field amplification through shear-driven MHD instabilities, including MRI-driven dynamos in disks and Tayler--Spruit-like mechanisms in radiative stellar interiors \cite{2005PhR...417....1B}. Tides may contribute indirectly by exciting gravity waves that transport angular momentum and help maintain differential rotation. Thus, while the nonlinear saturation and relaxation of the field remain uncertain, mass transfer supplies a plausible physical route to accretion-induced magnetic-field generation.}
Such dynamos are discussed in the literature, for example accretion disks are expected after the mergers of massive stars \cite{Schneider:2019} and white dwarfs \cite{GBerro:2012}, which could lead to a population of magnetized objects. Details of the dynamo process are not yet fully understood, so it is difficult to make strong predictions about the properties of the generated magnetic fields.  The problem of the relaxation of stellar magnetic fields and their final configuration has been studied, but only in idealized conditions \cite{Braithwaite2006}, showing that dynamo generated fields can in principle reach a stable configuration and be observable for long timescales.
However, the interaction of this process with rotation, stratification and mass loss has not been studied. We note that magnetic configurations similar to Plaskett's star (with a magnetic axis nearly perpendicular to the rotation axis) exist, and possibly more often in binary systems and/or rapidly rotating stars \cite[e.g.][]{2000A&A...359..213L,2019MNRAS.483.3127S,2013MNRAS.431.1513F}. This supports the {proposal} that some of these magnetic fields might have been generated via binary interactions, either via tidal interactions or mass transfer.

{The emergence of magnetic fields during mass transfer would significantly impact binary evolution. These magnetic fields can extract angular momentum \citep{2009MNRAS.392.1022U}, preventing the accreting star from reaching critical rotation velocities and thus sustaining efficient accretion. They also alter the stellar structure \citep[e.g.][]{2011A&A...534A.140C,2019MNRAS.487.3904M} and, if retained, may lead to the formation \citep{2008MNRAS.389L..66F} of magnetars (highly magnetised neutron stars whose origin is still debated). Since the majority of massive stars undergo mass transfer, this mechanism dramatically shapes our understanding of massive-star evolution.}

\newpage

\section*{\hypertarget{sec:methods}{Methods}}

\section*{Observations}\label{Sect:obs}

Archival high resolution ($R=65,000$), high signal-to-noise ratio (${\rm S/N}$ between 600-2000 per 1.8~km/s spectral pixel) ESPaDOnS spectroscopic observations of Plaskett's star were obtained from the Canada-France-Hawaii Telescope's (CFHT's) Canadian Astronomy Data Centre (CADC) archive (https://www.cadc-ccda.hia-iha.nrc-cnrc.gc.ca/en/cfht/). The reduced data, processed at the CFHT with the Upena pipeline, were downloaded and observations obtained on individual nights were co-added. The co-added spectra were normalized to the continuum using polynomial fits to each spectral order. Corrections to the heliocentric frame were applied. The details of the spectra are summarized in section 2 and Table 1 of \cite{2013MNRAS.428.1686G} and section 2.1 and Table 1 of \cite{2022MNRAS.512.1944G}.

\section*{Radial velocities}

For the narrow-line primary, a rich set of absorption and emission lines is available for RV measurement. We measured the RVs of various sets of lines (e.g., He\,{\sc i}\,$\lambda 4144$, He\,{\sc i}\,$\lambda 4173$,  N\,{\sc iii}\,$\lambda \lambda 5321, 5327$, and the N\,{\sc ii}\,$\lambda \lambda 5667, 5676, 	5680, 5686$ emission complex). 	The orbital solution for the primary does not depend on the choice of line within errors. The lowest $\chi^2$ is obtained for the emission complex in $5660 - 5710\,$\AA, dominated by  N\,{\sc ii} lines, presumably due to a lesser impact of variability caused by photospheric variability of the primary star.

The RV measurement for the broad-line secondary is much more challenging, both due to its broad line profiles, as well as due to the fact that almost all of its lines are contaminated by the primary star. Blended lines should be treated with more sophisticated methods such as spectral disentangling, which we perform below. However, one spectral line appears to belong to the secondary and to be essentially uncontaminated by the primary: the O\,{\sc iii}\,$\lambda 5592$ line. The core depth of this line is only 1\% of the continuum level, but is clearly visible owing to the high S/N of the spectra. 

A comparison between a few N\,{\sc ii} lines belonging to the primary and the O\,{\sc iii}\,$\lambda 5592$ line belonging to the secondary, as observed during RV extremes, is shown in Fig.\,\ref{fig:RVExt_Comp}. While the blue wing of the line appears rather stable, variable magnetospheric emission clearly impacts the red wing. {In Fig.\,\ref{fig:Dynspec} we also show a dynamic spectrum of the phased O\,{\sc iii}\,$\lambda 5592$ line. Each horizontal contribution to the 2D image represents the differential profile of the line at the indicated phase. It is evident from both of these illustrations that} the O\,{\sc iii}\,$\lambda 5592$ shows a small, but visually notable, anti-phase motion compared to the primary. 

\begin{figure*}
  \centering
\includegraphics[width=.41\textwidth]{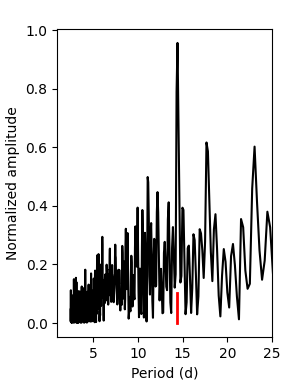}\hspace{0.5cm}\includegraphics[width=.404\textwidth]{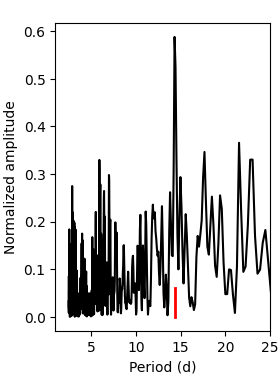}
    \caption{{Lomb-Scargle periodograms of the RVs of the narrow-line stripped primary (left) and broad-line magnetic secondary (right), respectively. Both timeseries recover the known \cite{Linder2008} period as the most significant signal.}} 
    \label{fig:LCP}
\end{figure*}

This fact strongly suggests that the two components are bound on the 14.4\,d orbit, and exhibit an extreme mass ratio. To verify this, we compute a Lomb-Scargle periodogram on the RVs of the secondary. We identify a clear peak corresponding to the well-established orbital period, as shown in Fig.\,\ref{fig:LCP}.

The RVs of the secondary were measured in the region 5584-5603\,\AA~to avoid contamination of the emission wings due to their different behavior compared with the line core, but the overall conclusions are independent of their exclusion. 
The phase-folded RVs of both components are shown  in Fig.\,\ref{fig:OrbitSolution}, and their relative amplitudes imply a mass ratio of $6.9^{1.2}_{-0.9}$.

{Fig.\,\ref{fig:RVExt_Comp} further demonstrates that the  O\,{\sc iii}\,$\lambda 5592$ line exhibits non-Keplerian variability. Such variability may originate in its magnetosphere, as well as in  spot activity that was previously invoked to explain photometric variability of the star \cite{Mahy2011}. Variability associated with these phenomena is expected to be either stochastic or bound to the rotational period of the star, estimated at 1.2\,d \cite{2022MNRAS.512.1944G}. While this 
 may well introduce additional scatter to our measurements which is not reflected by the statistical errors, the fact that this variability is not expected to follow the orbital period implies that it should not systematically affect our derived mass ratio, though it could result in underestimated errors. Our reported uncertainties on the mass ratio may therefore be underestimated. }

\section*{Spectral disentangling}
\label{subsec:specdissupp}

Briefly, the shift-and-add  algorithm is an iterative procedure that computes approximations for the disentangled spectra of the primary and secondary in the $j^{\rm th}$ iteration, $A_{\rm j}$ and $B_{\rm j}$, by shifting-and-subtracting the previous approximations $A_{\rm j-1}, B_{\rm j-1}$ from all observations, respectively, and then co-adding the residual observations in the respective frame-of-reference. To solve for the orbital elements simultaneously, the RVs of both components are assumed to trace a Keplerian orbit, and the goodness-of-fit for each set of orbital elements is evaluated via a $\chi^2$ minimization. {The initial guess for the secondary is assumed to be a flat (i.e., featureless) spectrum, }{while the initial primary spectrum is obtained by co-adding the observed spectra.}

\begin{figure*}
\centering
\begin{tabular}{ccc}
\includegraphics[width=0.3\textwidth]{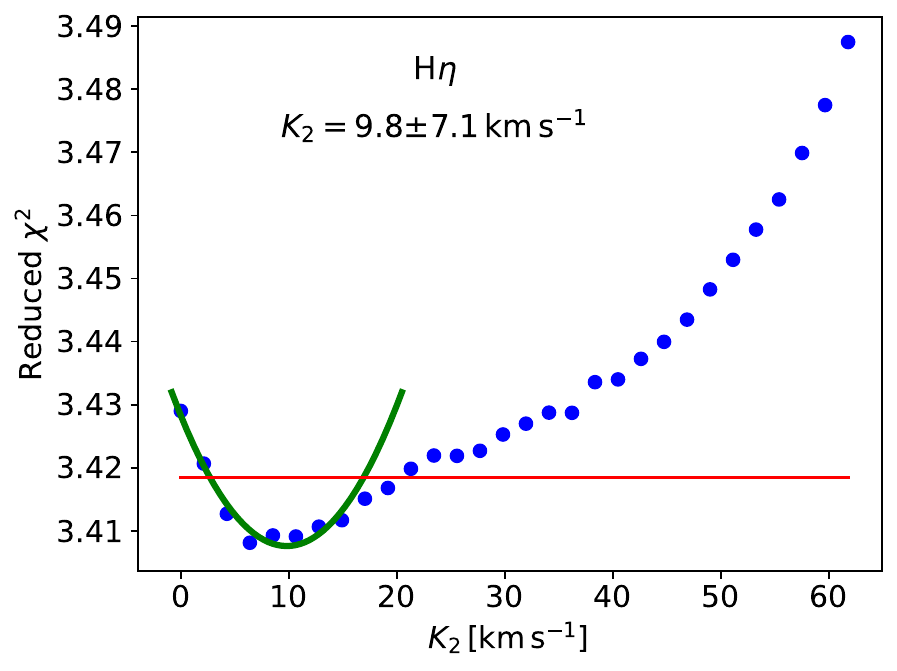} &
\includegraphics[width=0.3\textwidth]{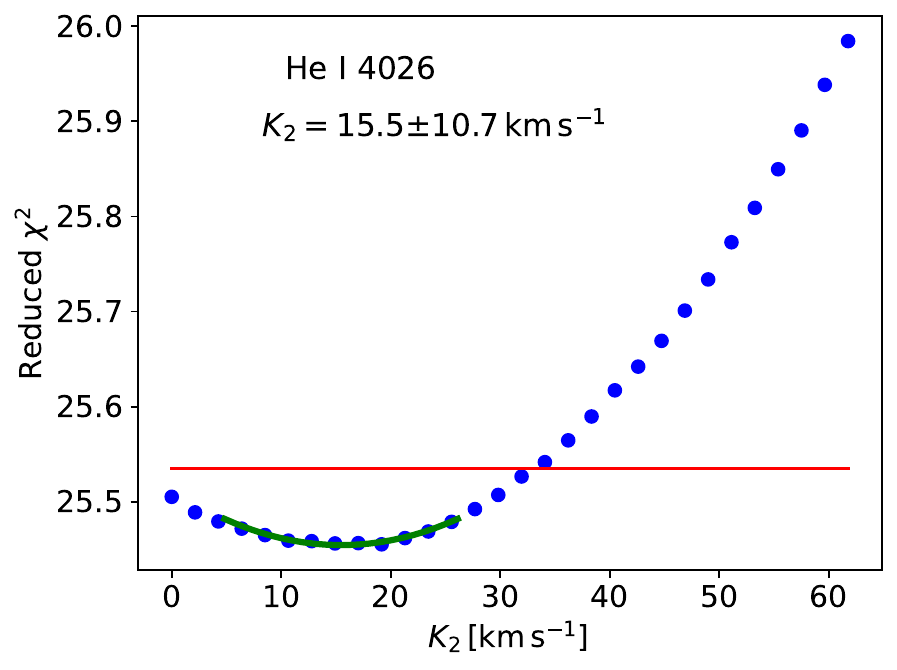} &
\includegraphics[width=0.3\textwidth]{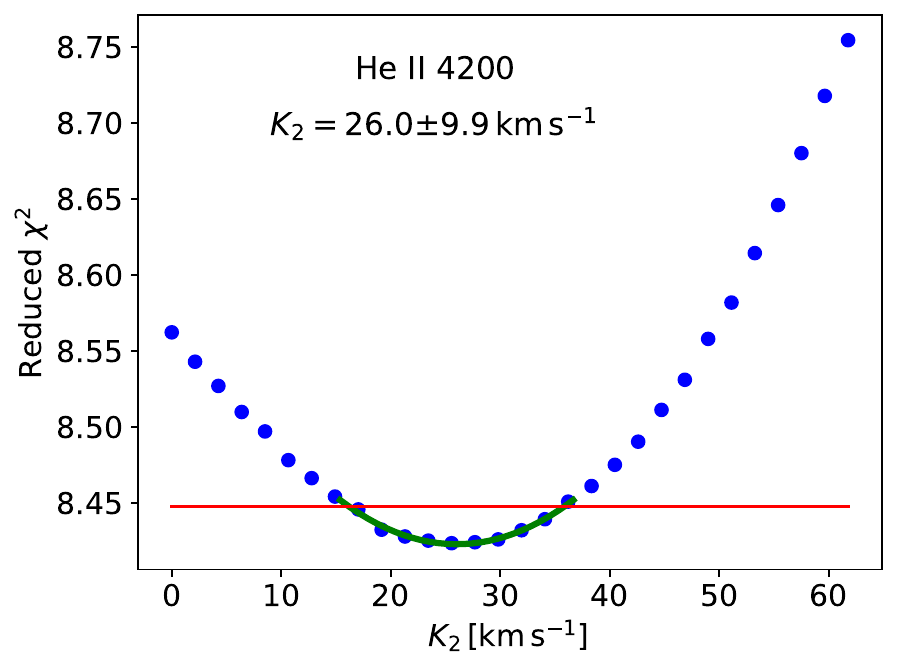}  \\
\includegraphics[width=0.294\textwidth]{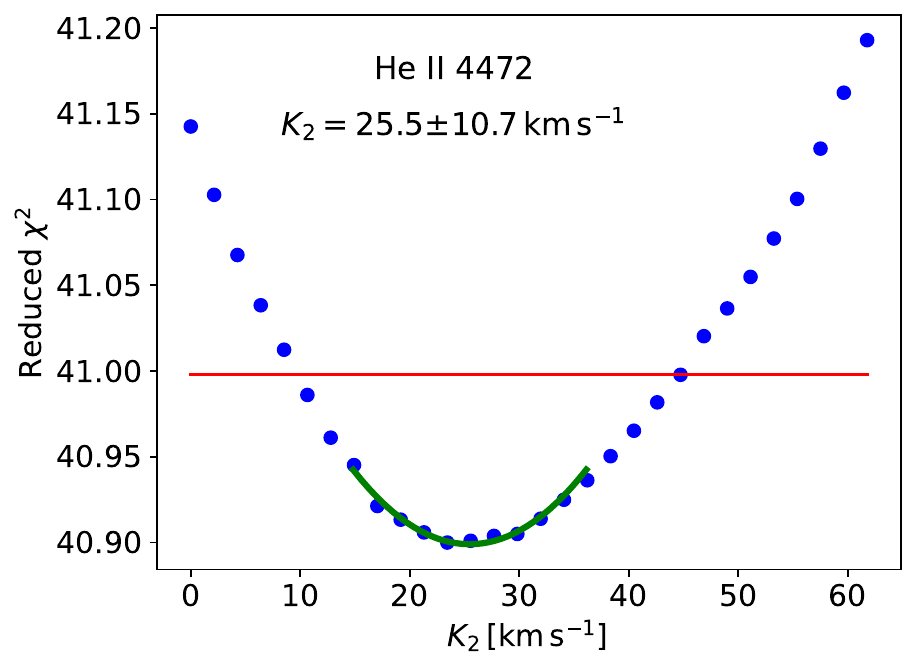} &
\includegraphics[width=0.3\textwidth]{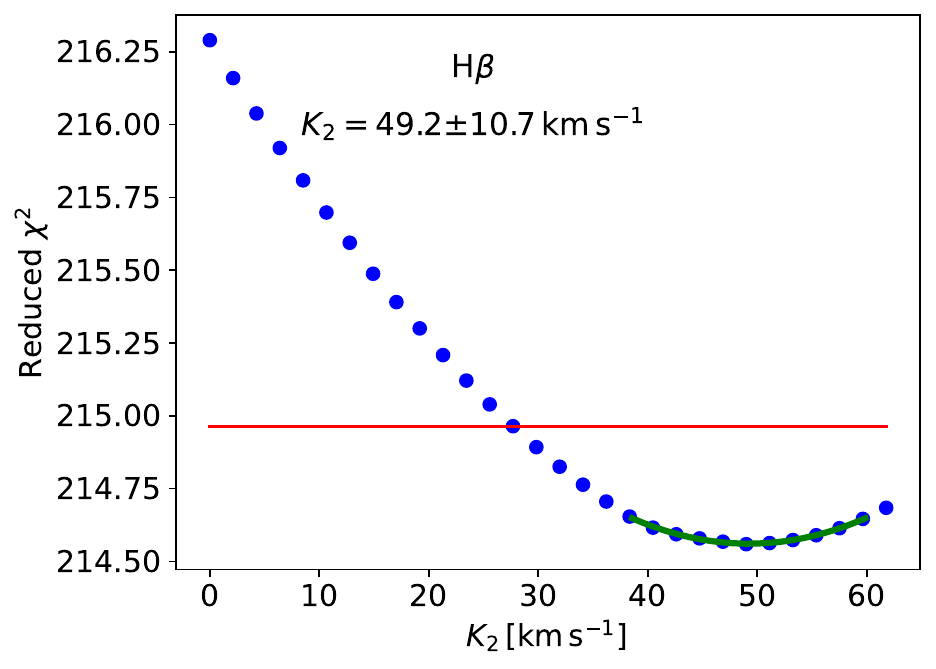} &
\includegraphics[width=0.3\textwidth]{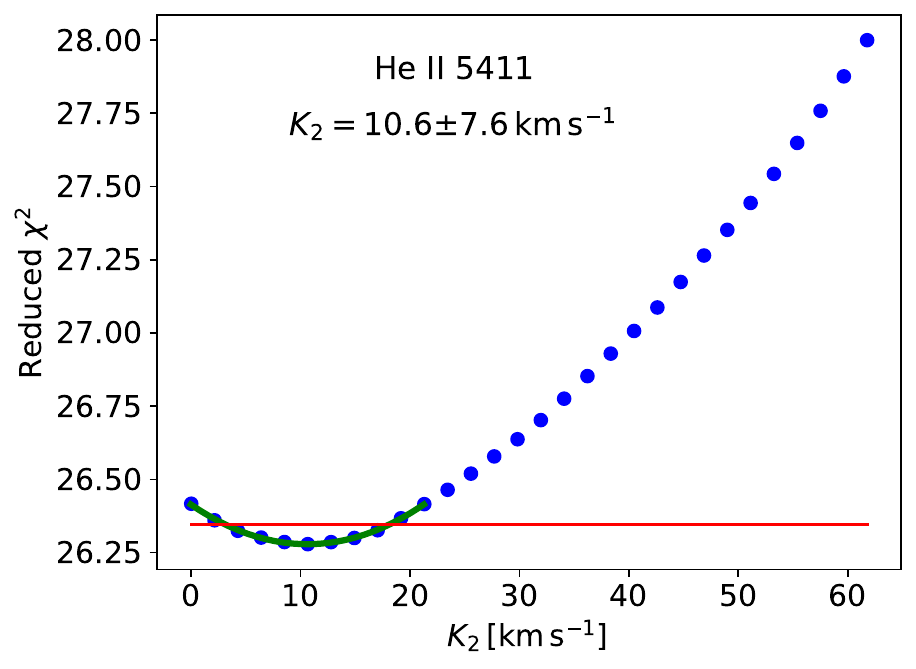}  \\
\end{tabular}
\caption{$\chi^2(K_2)$ from disentangling of the six lines shown in Fig.\,\ref{fig:DisLines}. The values and errors are derived by fitting a parabola to the minima, and are assumed to be symmetric for simplicity. }
\label{fig:chi2}
\end{figure*}

\begin{figure*}
  \centering
\includegraphics[width=\textwidth]{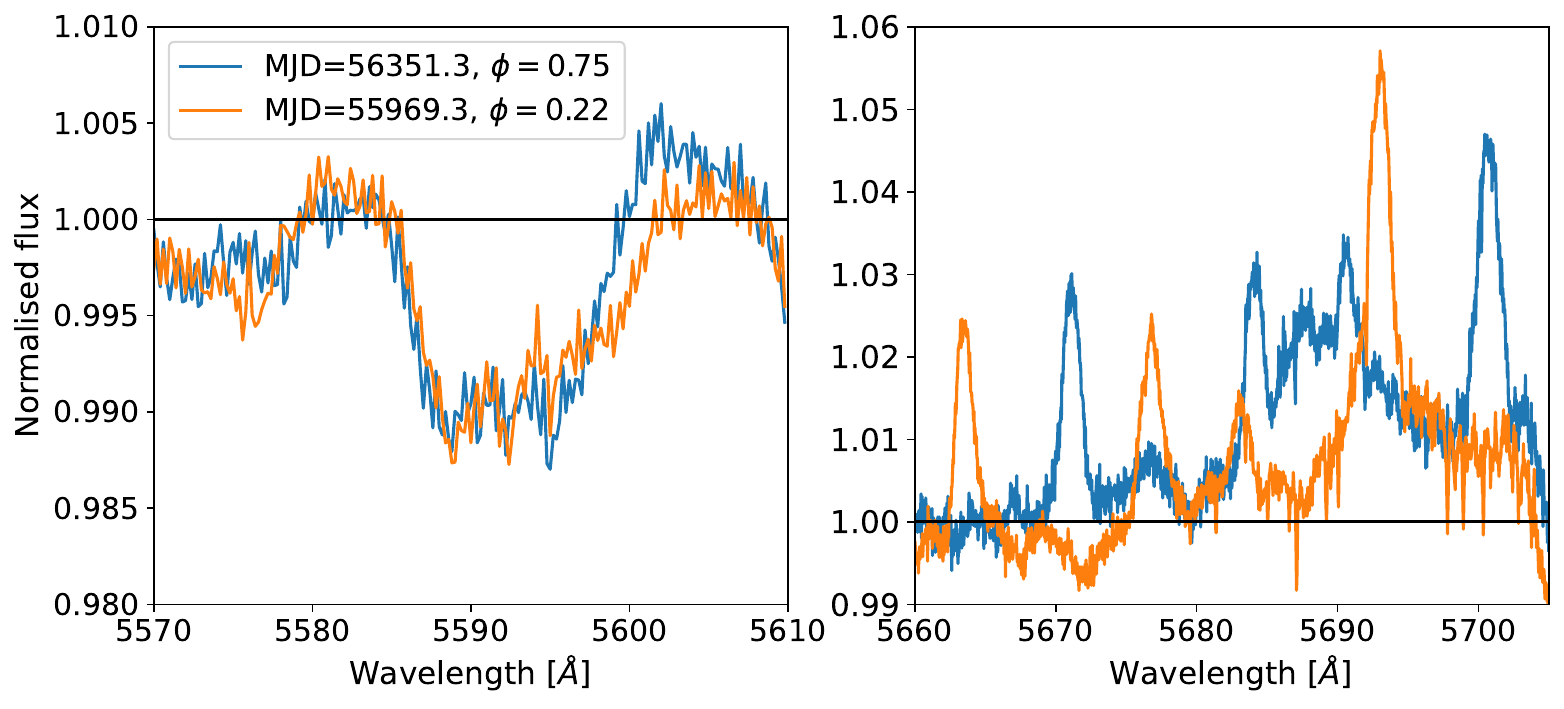}
    \caption{Comparison between two spectra at { radial velocity} extremes (see legend), zooming on the O\,{\sc iii}\,$\lambda 5592$ line (left panel), which originates solely in the magnetic secondary, and the emission complex in the range $5660 - 5710\,$\AA, which is dominated by the primary. The spectra in the left panel were binned at $\Delta \lambda = 0.2\,$\AA~for clarity. A slight anti-phase motion is apparent, suggestive of an extreme mass ratio between the components. {The narrow features seen in the {right} panel at phase $\phi = 0.22$ are telluric lines.} }
    \label{fig:RVExt_Comp}
\end{figure*}

\begin{figure}
  \centering
\includegraphics[width=.48\textwidth]{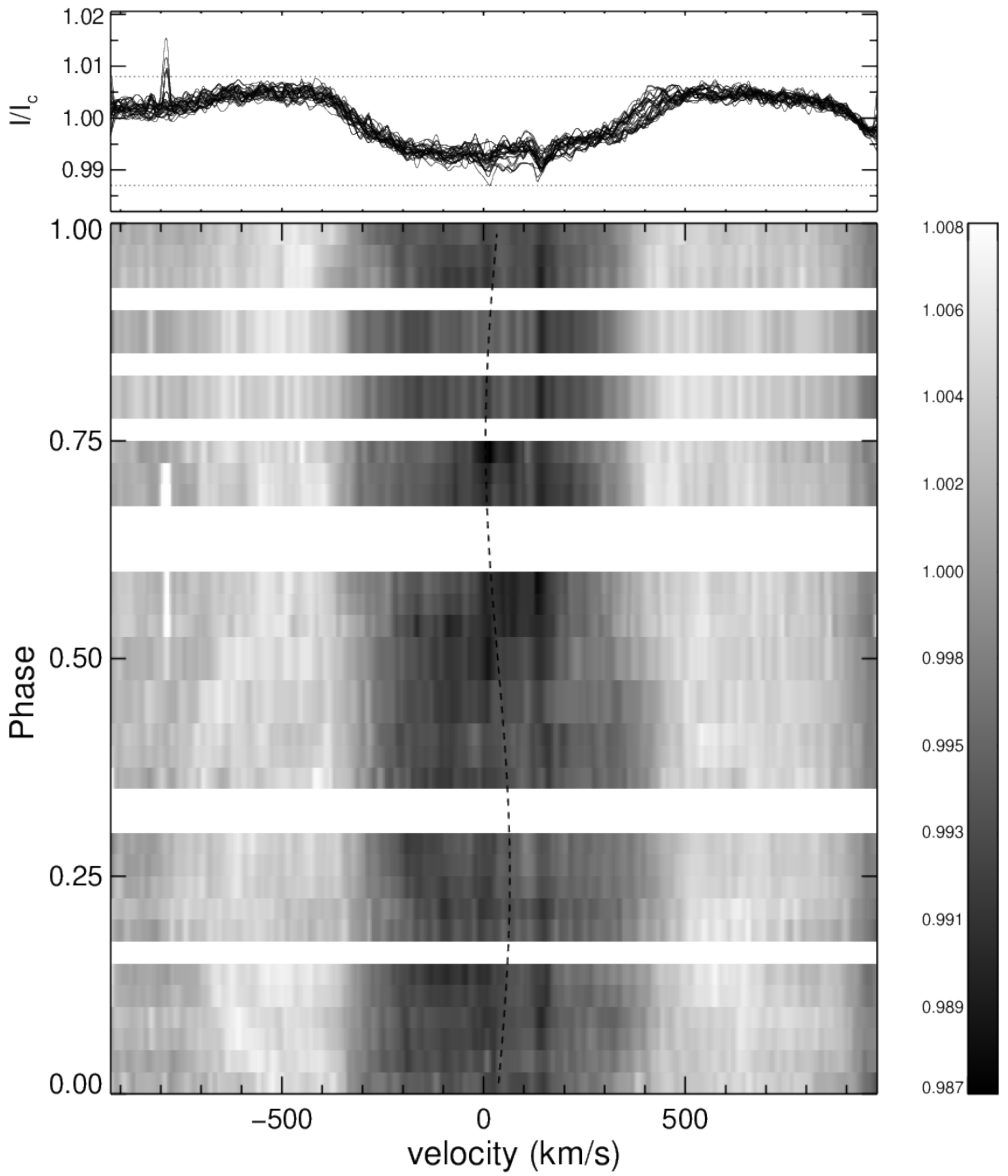}
    \caption{{{Dynamic spectrum of the O\,{\sc iii}\,$\lambda 5592$ line phased according to the 14.4~d RV period. The dashed curve represents the predicted RV variation (amplitude, phase, and period) due to the orbital motion inferred from the cross-correlation analysis. While the motion is small, it is still perceptible in the dynamic spectrum and is in excellent agreement with the expectation. The line profiles are shown overplotted in the upper panel.}}}
    \label{fig:Dynspec}
\end{figure}

Since all orbital elements are known to high precision except for $K_2$, in our disentangling we fix all the orbital elements to the values provided in Table\,\ref{tab:all_params}, and solve for $K_2$ by minimizing the reduced $\chi^2(K_2)$. Figure\,\ref{fig:DisLines} shows the disentangled spectra at RV extremes for six diagnostic lines, which were included in the analysis due to their relative strength and isolation. Figure\,\ref{fig:chi2} shows the corresponding reduced $\chi^2(K_2)$ maps. We note that, in most cases, $\chi^2 \gg 1$. This indicates that the non-Doppler variability in the lines is substantially larger than the S/N of the data. To obtain the 68\% confidence interval ($1\sigma$), we re-scale the errors such that $\chi^2=1$ when it reaches a minimum.

\section*{Physical parameters}

We use pre-computed spectral grids for OB-type stars  \cite{Hainich2019}. The grids span the 2D space of effective temperature -- surface gravity ($T_{\rm eff} - \log g$) and a solar abundance pattern \citep{Asplund2009}.  For the narrow-line primary, we compute additional models with varied abundances, since its spectrum implies strong deviations from solar abundances. We fix the microturbulent velocity to $v_{\rm mic} = 14\,$\kms, as assumed in the OB grid.

Given the highly complex spectra of both components, and the strong contamination by emission features originating from circumstellar and circumbinary material, we cannot expect to reproduce the spectra with standard spherical models. Instead, we aim to roughly reproduce the strengths and shapes of diagnostic features in the spectra. Errors are estimated by comparing the fits to nearby grid models.

As described in Results, subsection ``Spectral disentangling, we used several metal lines (whose abundances are expected to be solar) to fix the light contribution and effective temperature of the narrow-line primary. Figure\,\ref{fig:A_Metals} shows the disentangled spectrum of the narrow-line primary when assuming a light contribution of $l_1(V) = 55\%$ to a few PoWR grid models. While the line strengths depend on the grid parameters, the ionization balance (e.g., of Si\,{\sc iii} and Si\,{\sc iv}), and especially the Si\,{\sc iv}\,{$\lambda \lambda$ 4089, 4116} doublet,  implies $T_{\rm eff} = 31.0\pm1.0$. 
{  The primary's Balmer lines appear to suffer from substantial emission infilling, which makes the derivation of its $\log g$ challenging.  To estimate the primary's $\log g$, we resort to the helium lines (mainly He\,{\sc ii}) to estimate $\log g = 3.3 \pm 0.2\,[\cms]$. }

\begin{figure}
    \centering
    \includegraphics[width=\linewidth]{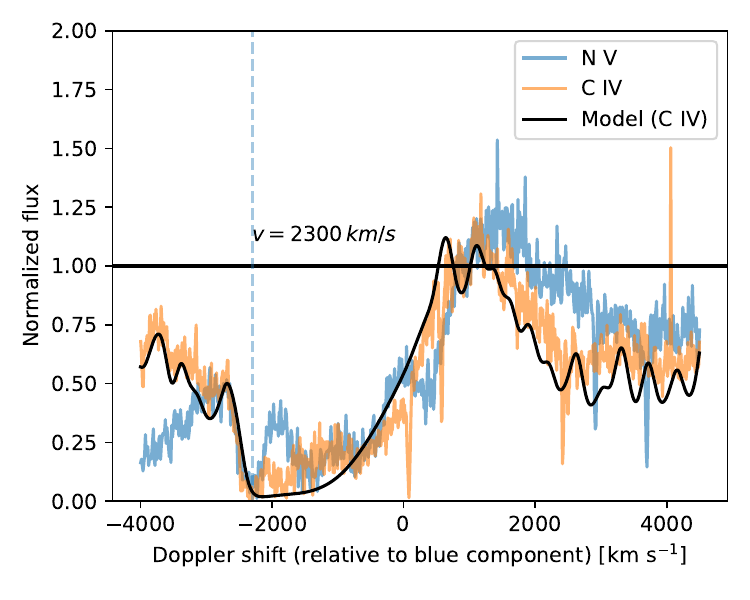}
    \caption{{Comparison between the P-Cygni resonance lines C\,{\sc iv}\,$\lambda \lambda 1548, 1551$ (orange) and N\,{\sc v}\,$\lambda \lambda 1239, 1243$ (blue) lines observed with IUE to the primary's model of the C\,{\sc iv} doublet, computed with $v_\infty = 2300\,$\kms~and $\log \dot{M} = -5.8\,[M_\odot\,{\rm yr}^{-1}]$. The IUE spectrum (ID: sp10689, JD = 2444570.91) was shifted to the rest wavelength based on our orbital solution for the primary.} }
    \label{fig:Pcyg}
\end{figure}

\begin{figure*}
\centering
\includegraphics[width=\textwidth]{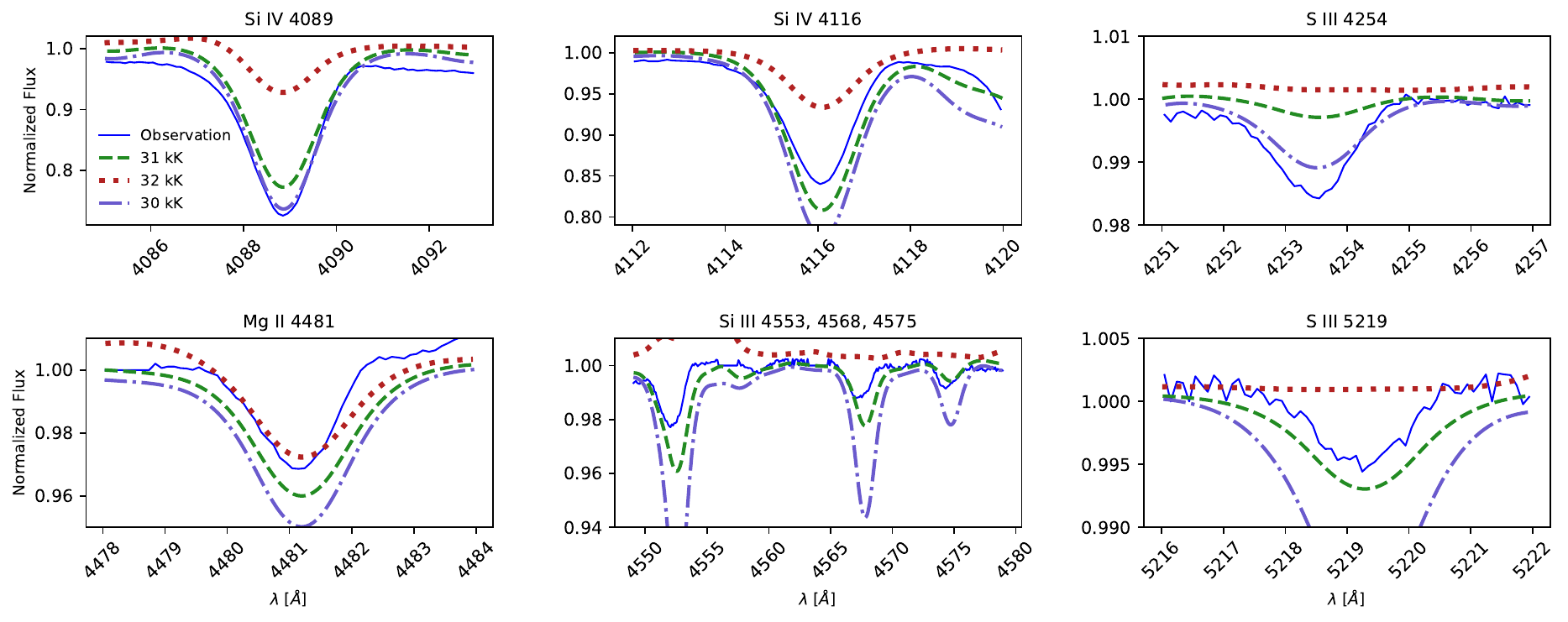} 
\caption{Observed disentangled spectrum of the narrow-line primary (blue solid line) compared to three grid PoWR models \citep{Hainich2019} with $\log g = 3.3\,[\cms]$ and varying effective temperatures of $T_{\rm eff} = 30, 31, 32\,$kK (see legend). The $T_{\rm eff} = 31\,$kK performed best on average at reproducing these diagnostic metal lines.  }
\label{fig:A_Metals}
\end{figure*}

 {The C\,{\sc iv}\,$\lambda \lambda 1548, 1551$ and N\,{\sc v}\,$\lambda \lambda 1239, 1243$ P-Cygni lines seen in the UV are saturated (Fig.\,\ref{fig:Pcyg}). Considering that both components are expected to contribute similarly in the UV, this implies that both components have saturated P-Cygni lines. However, since the magnetic star's lines are likely affected by the magnetosphere, we do not attempt to model its wind lines. Instead, we focus on the primary.} 
The saturated P-Cygni lines and emission in Balmer lines and He\,{\sc ii}\,$\lambda 4686$ imply that 
the stripped primary harbors a significant stellar wind.

We attempted to reproduce these features by adjusting the mass-loss rate $\dot{M}$ and terminal velocity $v_\infty$ of the primary, adopting a clumping factor of $D = 10$. The absorption troughs of UV P-Cygni lines (e.g., C\,{\sc iv}\,$\lambda \lambda 1548, 1551$) imply $v_\infty = 2300\pm100\,$\kms, while the emission strength of the He\,{\sc ii}\,$\lambda 4686$ emission feature implies $\log \dot{M} = -5.8\pm0.3$\,$[M_\odot\,{\rm yr}^{-1}]$ (Fig.\,\ref{fig:Pcyg}). 
However, we could not reproduce the line profiles of recombination lines in the optical, nor could we reproduce the infilling observed in the late Balmer members (e.g., H8 and H9). 
Given the mismatch of several line profiles, these results should  be taken with caution. A full analysis of the wind of the primary star is beyond the scope of this work.

\begin{figure*}
\centering
\includegraphics[width=\textwidth]{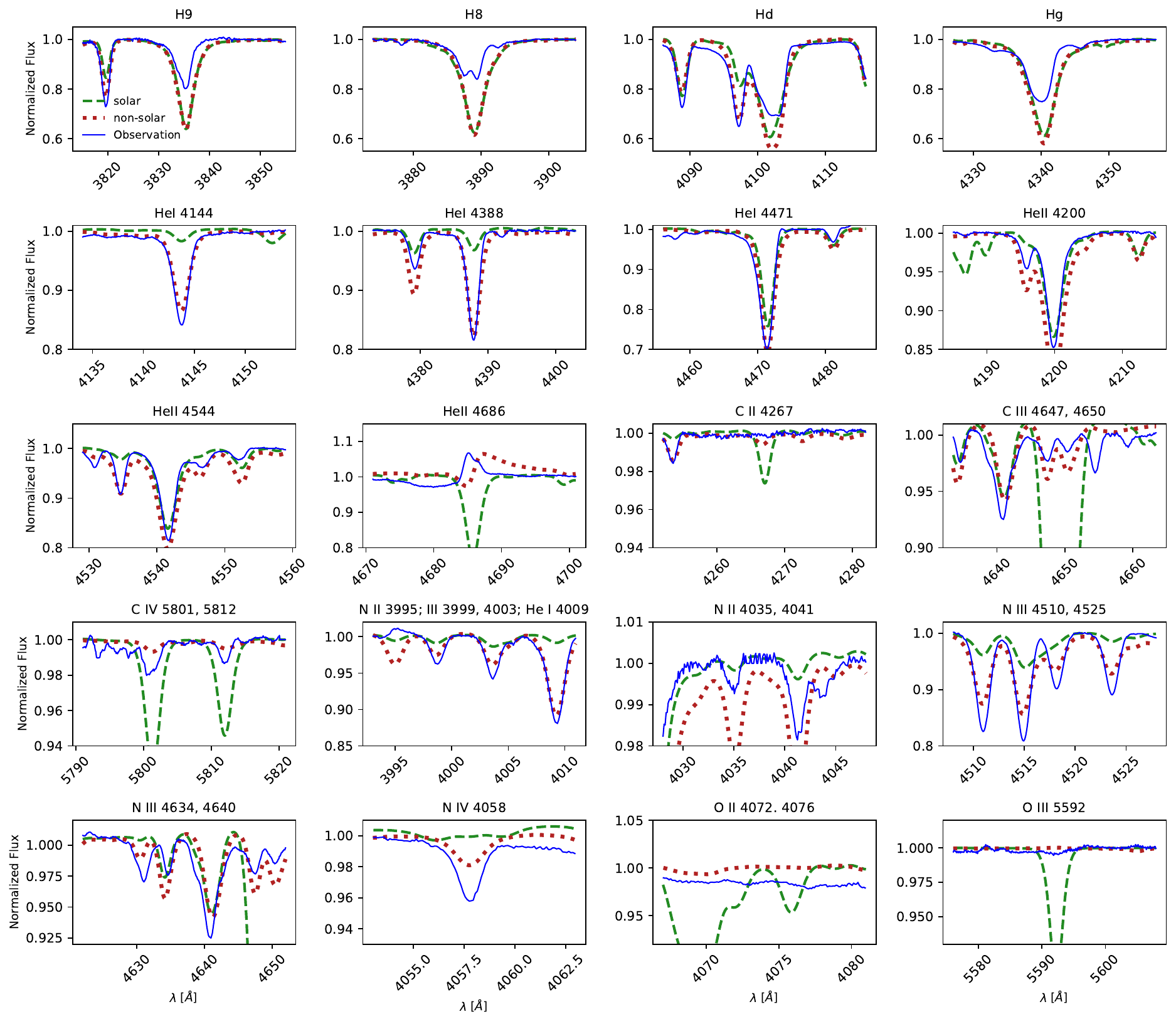} 
\caption{Observed disentangled spectrum of the narrow-line primary (blue solid line) compared to two models: the fiducial grid model with $T_{\rm eff} = 31\,$kK (Fig.\,\ref{fig:A_Metals}, and our adjusted model with similar parameters, but with an enhanced mass-loss rate and the non-solar abundance pattern given in Table\,\ref{tab:all_params}. }
\label{fig:FitA}
\end{figure*}

Figure\,\ref{fig:FitA} compares various observed spectral lines belonging to the primary with two models: a fiducial model with the parameters given above and a solar abundance pattern \cite{Asplund2009}, and a model with identical stellar parameters, but an enhanced N and He abundances and reduced C and O abundances (see Table\,\ref{tab:all_params}). 

\begin{figure*}
\centering
\includegraphics[width=\textwidth]{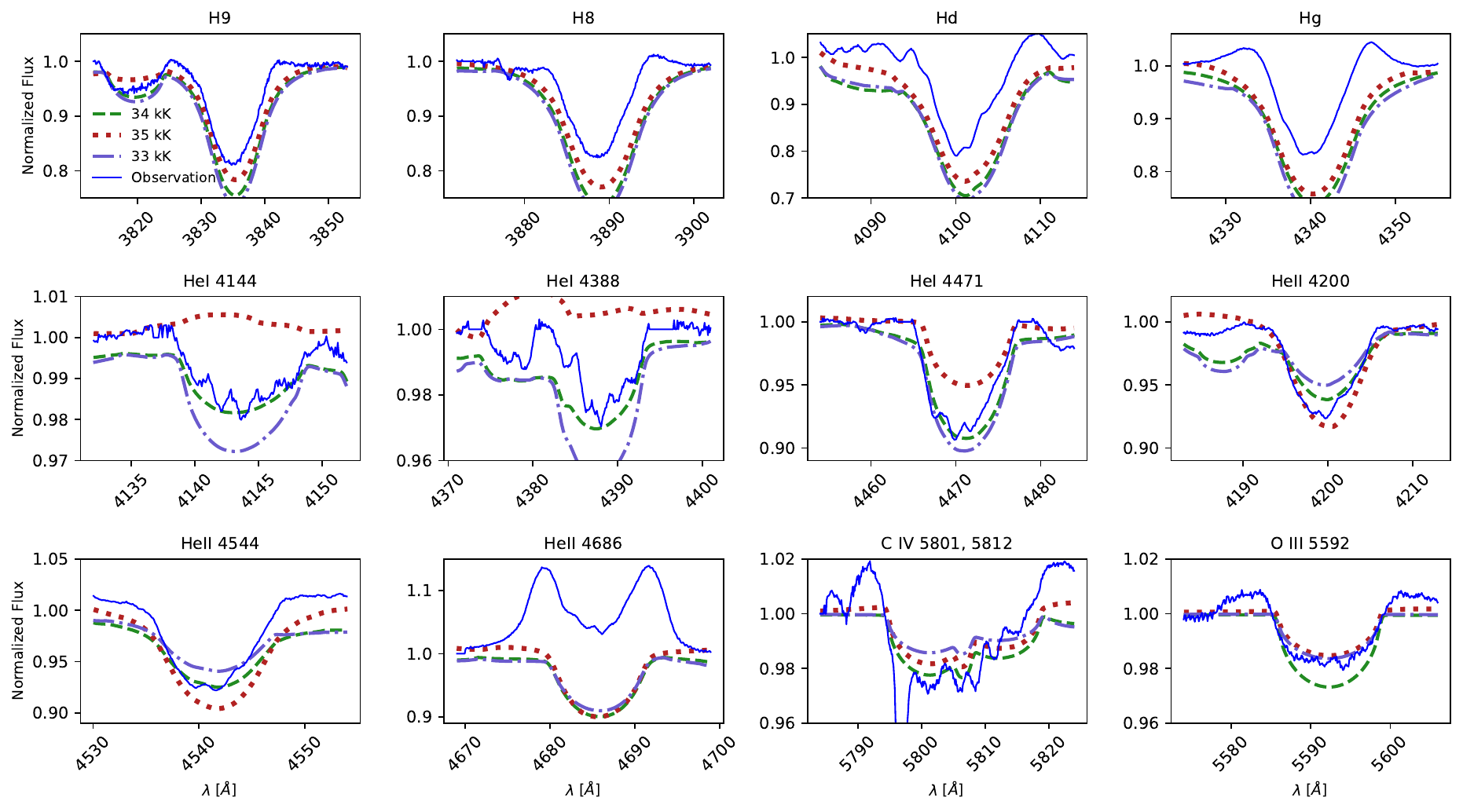} 
\caption{Observed disentangled spectrum of the broad-line secondary (blue solid line) compared to three grid PoWR models \citep{Hainich2019} with $\log g = 4.0\,[\cms]$ and varying effective temperatures of $T_{\rm eff} = 33, 34, 35\,$kK (see legend). The $T_{\rm eff} = 34\,$kK performs best in terms of reproducing the He\,{\sc i, ii} ionization balance. Some lines are heavily contaminated by emission.}
\label{fig:FitB}
\end{figure*}

For the broad-line secondary, very few spectral features are available, and the majority are strongly contaminated with emission. This specifically holds for all Balmer lines, which are the main $\log g$ diagnostics. 
In Fig.\,\ref{fig:FitB}, we show the disentangled spectrum of the broad-line secondary (scaled assuming a light contribution of $l_2(V) = 45\%$) compared with three models with $\log g = 4.0\,$[\cms] and varying $T_{\rm eff}$ values, focusing on diagnostic Balmer, He\,{\sc i, ii}, the C\,{\sc iv}\,$\lambda \lambda 5801, 5812$ doublet, and, importantly, the O\,{\sc iii}\, $\lambda 5592$ line.  The Balmer lines are systematically under-reproduced, and the most likely explanation for this is contamination by magnetospheric emission \citep{2022MNRAS.512.1944G}, which is not accounted for by the model.  For $\log g = 4.0\,[\cms]$, 
Fig.\,\ref{fig:FitB} shows that $T_{\rm eff, 2} = 34.0$~kK yields the best results when considering the He\,{\sc i, ii} lines. The best-fitting temperature would shift to $36\,$kK for $\log g = 4.3\,$[\cms], and to  $32$\,kK for  $\log g = 3.7\,$[\cms]. Hence, we estimate the secondary's effective temperature and its uncertainty at $T_{\rm eff, 2} = 34.0 \pm 2.0$~kK. We note that this temperature may be underestimated by up to about 2\,kK due to the rapid rotation of the star: the high $v \sin i$ value suggests that we are seeing the rotator close to edge-on. This exposes the observer primarily to the equator, which is cooler than the pole due to the Von Zeipel effect \cite{Abdul-Masih2023}.

For the narrow-line primary, we measure the projected rotational velocity $v \sin i$ and macroturbulent velocity $v_{\rm mac}$ using a mixture of Fourier analysis and goodness-of-fit against convolved profiles, implemented by the  {\scshape iacob-broad}  tool \cite{Simon-Diaz2007, Simon-Diaz2014, IACOBURL}. The analysis is performed on the relatively well isolated He\,{\sc i}\,$\lambda$4144 line, and yields $v \sin i_1 = 53.2 \pm 2.7$\,\kms~and $v_{\rm mac, 1} = 101.8 \pm 2.7\,$\kms. We convolve all PoWR models with corresponding profiles to account for this broadening, which yields a qualitatively satisfactory fit to the line shapes. For the broad-line secondary, the lines are too contaminated by non-stellar features to allow for a robust quantitative analysis. The line profiles suggest that rotation is the main source of broadening of the lines, and so we fix $v_{\rm mac, 2} = 0\,$\kms, and estimate $v \sin i_2 = 350\,$\kms~from the width of the lines. Fits to the primary's spectrum are shown in Fig.~\ref{fig:FitA}.

\begin{figure*}
  \centering
\includegraphics[width=.9\textwidth]{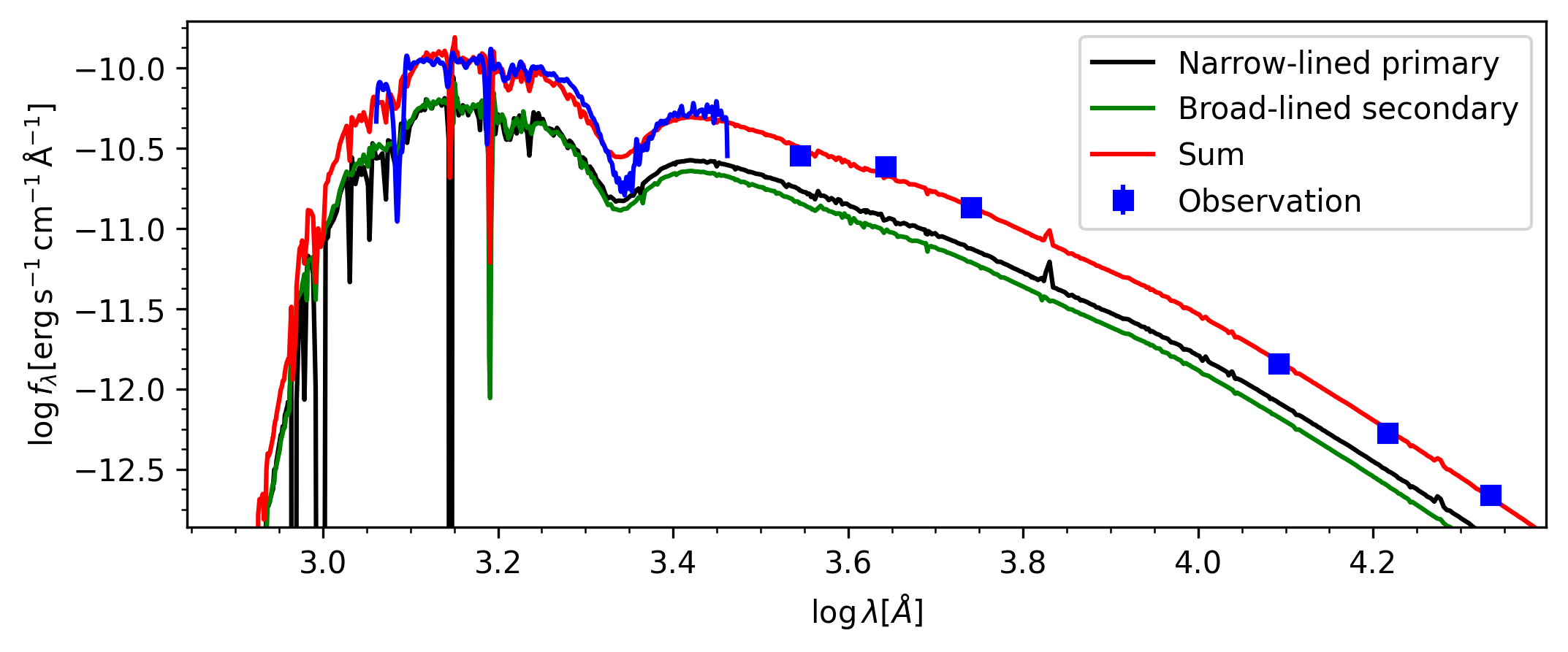}
    \caption{Observed photometry (blue squares; errors smaller than symbol sizes) compared to reddened synthetic spectral energy distribution of the binary (red), which is the sum of the primary (black) and secondary (green) { spectral energy distributions}. 
    } 
    \label{fig:SED}
\end{figure*}

To derive the luminosities $L$, we fit observed UBV \citep{Ducati2002} and JHK \cite{Cutri2003} photometry { and a co-added spectrum obtained with the International Ultraviolet Explorer (IUE)} \cite{Stickland1987}  to the sum of both synthetic spectral energy distributions (SEDs), fixing the light contribution of the primary in the visual band to $0.55$ as derived, and applying extinction as in \citep{Cardelli1989}, deriving the color excess $E_{B-V}$ and total-to-selective extinction ratio $R_V$. 
Figure\,\ref{fig:SED} shows the individual component SEDs and their sum compared to the observed photometry, from which the luminosities and radii are derived.

\subsection**{Constraints from photometry}


From the spectroscopic modelling, we can put reasonably stringent constraints on the projected rotation rates of the two components and we can infer the light ratios for each component that are needed to fit the spectra. Archival photometric data on the other hand show that the ellipsoidal variations {(if present)} {cannot be larger} {must be smaller} than $\Delta\mathrm{mag}=0.02-0.03$, {otherwise they would be clearly detectable} \citep[see e.g.,][]{Mahy2011}. Using the spectroscopic constraints in combination with the photometric constraints, we can rule out certain orbital configurations.

We use the \textsc{phoebe ii} code \citep[hereafter \textsc{phoebe}, ][]{Prsa2016} to model the light curve and rotation rates of each component under several different sets of assumptions. We begin by adopting the period from \cite{Linder2008} and the temperatures of the two components from the spectroscopic analysis above. As a starting set of assumptions, we assume that 1) the primary is in a semi-detached configuration and is filling its Roche Lobe, 2) the semi-detached component is tidally locked, 3) the secondary component is a rapid rotator and its axis of rotation is aligned with the orbital axis.  For the purposes of the modelling, we set the secondary to be critically rotating.  From the assumptions above, many parameters of the system are constrained so we only have 4 free parameters to vary: the mass ratio $q$, the mass of the secondary $M_2$, the radius of the secondary $R_2$, and the inclination $i$. With the assumption that it is filling its Roche lobe, the radius of the primary is constrained by the period of the system, the mass ratio and the mass of the primary. Table \ref{table:photometry} shows a summary of the different models that we computed, each of which are detailed below.

We begin the modelling assuming the best fit parameters from the spectroscopic fitting, namely $M_2 = 40~M_\odot$ and $R_2 = 10.4~R_\odot$ and adopting $i = 48^\circ$.  With this setup (mod 1 in Tab. \ref{table:photometry}{)}, the light ratios are far from the observed values {, the projected rotation rate of the primary ($v_1\sin i$) is too low} and the $\Delta\mathrm{mag}$ is too high. The combination of the $v_1\sin i$ and $\Delta\mathrm{mag}$ are additionally incompatible as {they require changing the inclination in opposite directions} {lowering the inclination to address the $\Delta\mathrm{mag}$ would result in a $v_1\sin i$ that is too low.}  Increasing the radius of the secondary (mod 2) can help with the light ratios, and increasing the mass of the secondary (and thus the primary through the fixed $q$, mod 3) can increase the $v_1\sin i$, but the value of $\Delta\mathrm{mag}$ is still much too far from the observed range.  Adjusting the mass ratio (mod 4) has a slight effect on $\Delta\mathrm{mag}$, but not enough to make a significant difference. 

Starting again from the initial setup but relaxing the assumption that the primary is filling its Roche lobe and instead setting its radius to 60\% of the Roche lobe radius ($\nicefrac{R_1}{R_{RL}}$ = 0.6, mod 5) results in a much lower $\Delta\mathrm{mag}$, but the $v_1\sin i$ is in disagreement with the observations.  Given that the primary is no longer filling its Roche lobe, it no longer needs to be tidally locked, so we can adjust the synchronicity parameter ($F_1$).  Assuming that the system transferred mass in the past, and has since shrunk within its Roche lobe, it is reasonable to assume that as it shrunk, its rotation rate increased.  Approximating the star as a solid uniform sphere and conserving angular momentum, we can adjust $F_1$ so that the star matches the angular momentum that it had when filling its Roche lobe (mod 6), allowing us to raise the $v_1\sin i$ to acceptable values.  While the $v_1\sin i$ is slightly higher than the observed value, a more appropriate moment of inertia describing the star would likely fix this disagreement.  Overall, we reach a very good agreement between all of the observed and theoretical parameters in this configuration. Note, however, that $v_{2, \mathrm{max}}\sin i$ is quite a bit higher than the observed value implying that the secondary is not rotating at critical velocity.  Given that this component is magnetic, it is possible that magnetic braking has begun to slow the rotation, reducing it to the observed value.

    \begin{table*}[t] \label{table:photometry}
    \caption{Results of the PHOEBE {code} modelling.  Values {of parameters} that do not fit the observables are highlighted in red. {$F_1$ is the  synchronicity parameter.}}
    \centering 
    \setlength{\tabcolsep}{2pt}
    \begin{tabular}{ccccccc|ccccccc}
    \hline\hline
    \multicolumn{7}{c}{Inputs} & \multicolumn{7}{c}{Outputs} \\
    & $q$ & $\nicefrac{R_1}{R_{RL}}$ & $R_2$ & $M_2$ & $i$ & $F_1$ & $R_1$ & $L_1$ & $L_2$ & $v_\mathrm{2, crit}$ & $v_{2,\mathrm{max}} \sin{i}$ & $v_1 \sin{i}$ & $\Delta\mathrm{mag} $\\
    & $\nicefrac{M_2}{M_1}$ & & [\rsol] & [\msol] & [$^\circ$] & & {[\rsol]} & $\nicefrac{}{L_\mathrm{tot}}$ & $\nicefrac{}{L_\mathrm{tot}}$ & [\kms] & [\kms] & [\kms] & [mag]\\

    \hline  


    mod 1 & 6.9 & 1.00 & 10.40 & 40 & 48.0 & 1.00 & 20.38 & \red{0.780} & \red{0.220} & 773.23 & 574.62 & 53.23 & \red{0.132} \\
    mod 2 & 6.9 & 1.00 & 17.00 & 40 & 48.0 & 1.00 & 20.38 & 0.571 & 0.429 & 604.78 & 449.44 & 53.23 & \red{0.099} \\
    mod 3 & 6.9 & 1.00 & 17.00 & 60 & 48.0 & 1.00 & 23.33 & \red{0.635} & \red{0.365} & 740.70 & 550.45 & 60.94 & \red{0.108} \\
    mod 4 & 10.0 & 1.00 & 17.00 & 60 & 48.0 & 1.00 & 20.67 & 0.578 & 0.422 & 740.70 & 550.45 & 53.99 & \red{0.102} \\
    mod 5 & 6.9 & 0.60 & 10.40 & 40 & 48.0 & 1.00 & 12.23 & 0.541 & 0.459 & 773.23 & 574.62 & \red{31.94} & 0.016 \\
    mod 6 & 6.9 & 0.60 & 10.40 & 40 & 48.0 & 2.78 & 12.23 & 0.545 & 0.455 & 773.23 & 574.62 & 88.73 & 0.019 \\


    \hline
    \end{tabular}
    \end{table*}

\section*{MESA simulation setup}\label{mesa_setup}
{
We make use of version \texttt{r24.08.01} of the \texttt{MESA} code together with version \texttt{x86\_64-linux-24.7.1} of the \texttt{MESA} SDK. This version of \texttt{MESA} is described in its various instrument papers \cite{Paxton+2011,Paxton2013,Paxton+2015,Paxton2018,Paxton2019,Jermyn2023}. Initial composition is based on the GAL setup described by \cite{Brott+2011} and we make use of custom OPAL opacity tables \cite{IglesiasRogers1996} that match this specific choice of abundances. Our wind prescription also follows that of \cite{Brott+2011}, which within the range covered by our stellar tracks is a combination of the prescriptions of \cite{Vink2001} and that of \cite{Hamann1995}.}

{Stellar models are computed assuming non-rotating stars. Convection is modeled based on mixing length theory \cite{Bohm-Vitense1958} as described by \cite{CoxGiuli1968} with an efficiency parameter of $\alpha_\text{MLT}=1.5$. Convective boundaries are determined using the Ledoux criterion \cite{Ledoux1947} and we include semiconvective mixing as described by \cite{Langer1983} with an efficiency parameter $\alpha_\text{sc}=1$. Thermohaline mixing also plays an important role in mixing the helium rich material passed to the accretor with the rest of its hydrogen envelope, this process is modeled following the description of \cite{Kippenhahn1980} with an efficiency parameter of unity. We make use of the \texttt{basic.net} nuclear network of \texttt{MESA}, which includes the isotopes $^1$H, $^3$He, $^4$He, $^{12}$C, $^{14}$N, $^{16}$O, $^{20}$Ne and $^{24}$Mg. This reaction network is slightly modified to include $^{40}$Ca and $^{56}$Fe, both of which do not react with other elements in the network. In this setup iron is included to scale wind prescriptions in non-Galactic environments, and thus plays no role for the simulations in this work. Calcium is used as an inert element to store all mass associated with isotopes not present in the nuclear network. Nuclear reaction rates are given by a combination of results from JINA REACLIB \cite{Cyburt2010} and NACRE \cite{Angulo1999}, with screening determined following \cite{Chugunov2007}. \texttt{MESA} makes use of multiple sources for its equation of state, for the evolutionary phases covered by our calculations we rely on \texttt{FreeEOS} \cite{Irwin2004}.}

{Mass transfer is modeled using the \texttt{contact} scheme, which sets the mass transfer rate implicitly such that donors in semidetached systems remain just inside their Roche lobes (using the approximation for Roche lobe radii of \cite{Eggleton1983}), or that both components stay within the same equipotential in the case of a contact binary \cite{Marchant2016}. Although we include a treatment for contact systems note that the simulation shown in Figure \ref{fig:HR_evo} does not undergo a contact phase. Orbital evolution is treated based on angular momentum conservation and the assumption of a circular orbit \cite{Paxton+2015}. We assume conservative mass transfer, such that the only forms of mass and angular momentum loss from the system are stellar winds. We assume stellar winds remove a specific angular momentum equal to that of the corresponding mass-losing component.}

{As an initial search for solutions matching the present day mass ratio and luminosities of the components did not succeed, we performed a grid of simulations where the initial total mass, the initial mass ratio and the present day mass ratio were taken as parameters. Using these and the present day period of Plaskett's Star initial periods were determined using Equation (\ref{equ:orbevo}) which assumes conservative mass transfer. Our grid considered total initial masses between $30M_\odot$ and $50M_\odot$ in steps of $2.5M_\odot$, initial mass ratios $q_\text{i}$ between $0.6$ and $0.99$ in steps of $0.01$ and present day mass ratios $q$ between $3.5$ and $7$ in steps of $0.25$. This amounts to 5400 simulations.}

{Our assumption of conservative mass transfer serves as an optimistic physical assumption to try and match the observed mass ratio. Mass transfer efficiency remains as a significant uncertainty in binary evolution models, and particularly, the evolution of a mass transferring system after the accretor reaches critical rotation is uncertain (see discussions in \cite{Langer2012} and \cite{MarchantBodensteiner2024}). Recent observations of post-mass transfer systems with extreme mass ratios suggest nearly conservative mass transfer \cite{Lechien2025,Picco2026}, which would indicate that accretion can proceed independent of rotation. Moreover, existing grids of massive binary simulations that include {inefficient} mass transfer associated with rotation struggle to produce post mass-transfer mass ratios larger than $2$ \cite{Schurmann2024}.}



\section*{Calculation of the critical field}\label{bc_calculation}

{In accreting binaries, whether the accretion flow is magnetically channeled or directly overwhelms the surface field depends on the competition between the magnetic pressure at the stellar surface,
$P_{\rm mag}=B^2/8\pi$, and the ram pressure of the inflowing material,
$P_{\rm ram}=\rho v^2$. Neglecting thermal and radiation contributions,
the critical field strength $B_c$ -- equivalently, the field for which
the magnetospheric radius equals the stellar radius -- follows from the
condition $P_{\rm mag}=P_{\rm ram}$.}

The density at impact is set by the accretion rate, $\rho_{\rm impact} = \dot{M}/(v_{\rm impact} A_{\rm impact})$, where $\dot{M}$ is the mass transfer rate, $A_{\rm impact}$ the effective area of the stream, and $v_{\rm impact}$ the free-fall velocity at the stellar surface, $v_{\rm impact} = \sqrt{2GM/R}$, with $R$ the stellar radius. Substituting into $P_{\rm ram}$ gives $P_{\rm ram}=\dot{M} v_{\rm impact}/A_{\rm impact}$. Balancing the two pressures yields $B_c^2 = 8\pi \dot{M} v_{\rm impact}/A_{\rm impact} = 8\pi \dot{M} \sqrt{2GM/R}/A_{\rm impact}$. For simplicity, we parameterize the impact area as $A_{\rm impact} = \alpha R^2$, where $\alpha$ encodes the geometry of the stream. We recover the critical field in Eq.~\ref{eq:critical}:
\begin{equation}
B_c = \sqrt{\frac{8\pi \dot{M} \sqrt{2GM}}{\alpha R^{5/2}}}.
\end{equation}
This expression makes explicit the scaling of $B_c$ with accretion rate, stellar parameters, and flow collimation: stronger fields are required for higher $\dot{M}$, more compact stars, or narrower streams (smaller $\alpha$). The derivation assumes steady accretion, neglects thermal pressure, reconnection, and magnetic diffusion, and treats the surface field as uniform. { Within these limitations, the above expression provides a simple estimate of the magnetic field strength required to magnetically channel the accretion flow at the stellar surface. We stress that $B_c$ is not the field required to avoid burial: as discussed in Discussion, subsection ``Origin of the magnetic field, and implications on stellar evolution, even channeled accretion buries the field once the accreted mass exceeds a small critical value.}

{For completeness, we also estimate the field strength
required to resist burial. A field can only avoid being screened if its
magnetic pressure is comparable to the hydrostatic pressure at the base
of the accreted layer. For an accreted envelope of mass $\Delta M$ on a
star of mass $M$ and radius $R$, this pressure is of order
\begin{equation}
P_{\rm base} = \frac{G M \Delta M}{4\pi R^4},
\end{equation}
yielding a burial-resisting field
\begin{equation}
B_{\rm bury} = \sqrt{\frac{2 G M \Delta M}{R^4}}
=3\times10^{7}\,{\rm G}
\left(\frac{M}{20\,M_\odot}\right)^{1/2}
\left(\frac{\Delta M}{10\,M_\odot}\right)^{1/2}
\left(\frac{R}{9\,R_\odot}\right)^{-2}.
\end{equation}
Since $B_{\rm bury} \gg B_c$ for any relevant parameters, all fields in
the observed range of magnetic OB stars are expected to be buried by the mass transfer
inferred for Plaskett's star.}

{\section*{{On the discrepancy between the observed mass ratio and that predicted by evolutionary modeling}\label{sec:discrepancy}}

{While our favoured model reproduces the broader characteristics of the system, it reaches a final mass ratio of $q_{\rm model}=4$. This value is roughly 3$\sigma$ below the measured value of $q_{\rm obs}=6.9^{+1.2}_{-0.9}$ from the orbit. We conclude, based on the large grid of models that we have explored, that this mismatch is difficult to avoid within the mass range required to reproduce the observed luminosities of the system, and highlights a significant challenge for current binary evolution models of stripped stars. From our grid of simulations we can determine the maximum mass ratio we can reach at detachment while being consistent with the present day orbital period. This is shown in Figure \ref{fig:qdet}, which highlights how higher mass systems further diverge from the determined mass ratio.}

\begin{figure}
  \centering
\hspace{-0.25cm}
\includegraphics[width=9cm]{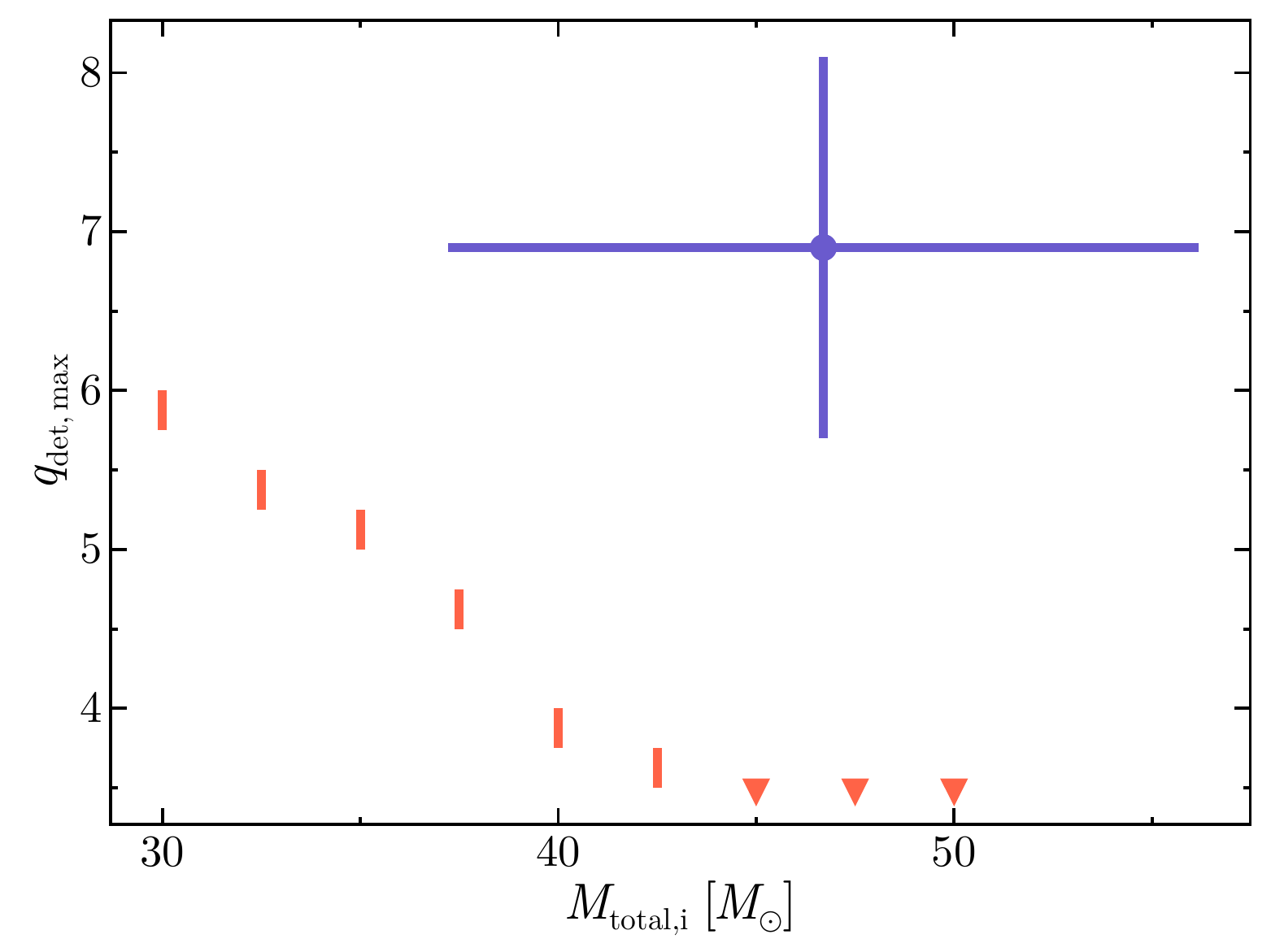}
    \caption{{Maximum mass ratios that are reached in our grid of simulations while remaining consistent with the present day orbital period. Results are shown as a function of total initial mass of the simulation, with bars representing the uncertainty given by the resolution of our grid. For total masses $\geq 45\,M_\odot$ no solutions are found down to the lower limit in the present day mass ratio considered on our grid ($q=3.5$), so only upper limits are shown. Plaskett's Star is shown in purple considering its present-day mass (with errors added in quadrature) and mass ratio as determined from the RV analysis. This model exploration clearly demonstrates the failure of theory to reproduce the observed mass ratio at any mass, with the discrepancy being particularly stark at masses comparable to those inferred for the system.}} 
    \label{fig:qdet}
\end{figure}

{Various other simulations from our grid are presented in Figure \ref{fig:HR_more}, showing systems that reach more extreme mass ratios but with a lower total mass, or higher mass systems that can only reach much lower mass ratios at detachment. In general, simulations that reach mass ratios comparable to that of Plaskett's Star have significantly underluminous stripped stars and are significantly undermassive, while higher mass systems have much lower mass ratios at detachment and are overluminous. Figure \ref{fig:HR_more} also shows a system with the same total initial mass as the one presented in Figure \ref{fig:HR_evo} but with a larger mass ratio at detachment of $q= 5$. The individual luminosities of each component in this simulation show a larger discrepancy, with the accretor being overluminous and the donor being underluminous. Despite the mismatch between observations and simulations, the nearly circular orbit and rapid rotation of the secondary are strong evidence of a stable mass transfer phase. In contrast, common envelope evolution is not expected to be a viable scenario, as it results in highly non-conservative evolution which leads to less extreme mass ratios.}

\begin{figure}
  \centering
\hspace{-0.25cm}
\includegraphics[width=9cm]{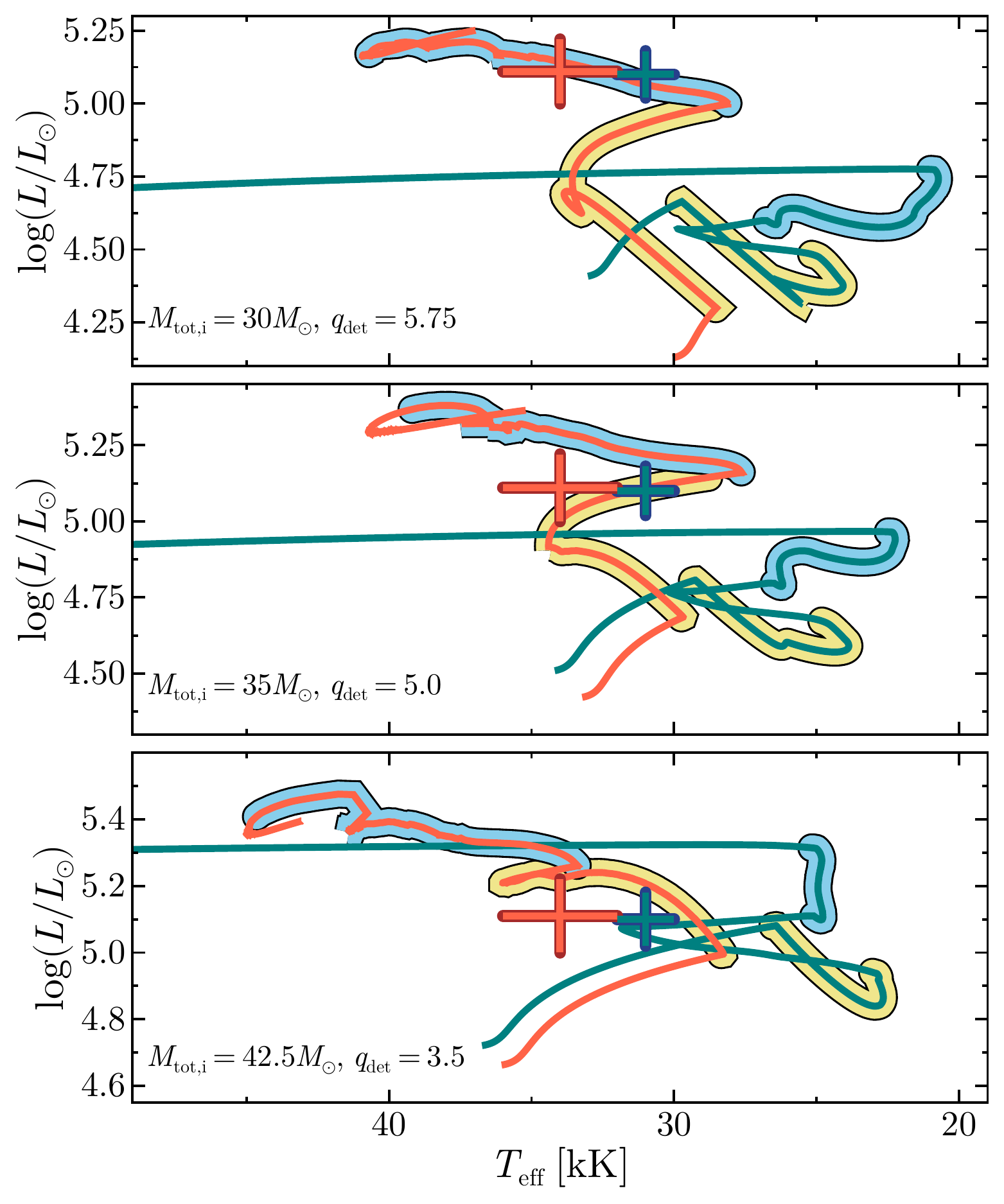}
    \caption{{Different examples from our grid of simulations that reach the most extreme mass ratio at a given total initial mass, while remaining consistent with the observed orbital period. $q_\text{det}$ represents the present day mass ratio used to determine the initial orbital period with Equation \ref{equ:orbevo}. (Top) System with a total mass of $30M_\odot$, initial mass ratio of $0.8$ and initial period of $1.9$ days, resulting in a $4.2M_\odot$ stripped star with a $24.2M_\odot$ companion in a $14.7$ day period orbit. (Middle) System with a total mass of $35M_\odot$, initial mass ratio of $0.93$ and initial period of $2.5$ days, resulting in a $5.4M_\odot$ stripped star with a $27.1M_\odot$ companion in a $14.6$ day period orbit. (Bottom) System with a total mass of $42.5M_\odot$, initial mass ratio of $0.95$ and initial period of $4.8$ days, resulting in an $8.6M_\odot$ stripped star with a $30.3M_\odot$ companion in a $15$ day period orbit.}
    } 
    \label{fig:HR_more}
\end{figure}

{Similar discrepancies were encountered in the modelling of the stripped-star system HR 6819 \cite{Picco2026}, potentially pointing to missing physics that would {enable more efficient stripping of massive stars} {result in less massive helium cores} or variations in orbital evolution. On the other hand, the spectra exhibit several sources of non-Keplerian variability (e.g., magnetospheric features, wind variability, and potentially circumstellar material), which could introduce systematic uncertainties into the inferred RV amplitudes and therefore the derived mass ratio.}

\section*{On the {fundamental conflict} with the previously reported orbit}\label{sec:mismatch_orbit}

{
As discussed in Results, subsection ``Radial velocities and orbital analysis, Plaskett's star was subject to a previous orbital solution and spectral analysis by \cite{Linder2008}. While the parameters derived for the primary agree between the two studies, their derived RV amplitude for the secondary of $K_2 = 192\,$\kms~is a factor 6  larger than the value derived here (30\,\kms). Inspection of the slowly-moving O\,{\sc iii} $\lambda5592$ line alone (to which the authors did not have access) readily suffices to reject the historic solution.  However, one may still wonder as to the origin of the mismatch with our solution, given the similar analysis methodology implemented in both studies.
}

{
While \cite{Linder2008} also used spectral disentangling {(employing the algorithm of \cite{Gonzalez2006})}, their analysis algorithm differs in one important way. To disentangle the spectra, the authors made use of a first set of measured RVs for both components. Those RVs were derived using "comb" functions for both components, defined to be null everywhere except at line centers, where it is unity. These functions were then cross-correlated with the data to yield a first set of RVs.}

{
This first step likely resulted in highly biased RV measurements, for two key reasons. First, all spectral lines used by these authors are blended and contaminated with emission, likely yielding inaccurate RVs. The authors then proceeded to derive the disentangled spectra, which were then used to derive the RVs, iteratively. However, using wrongly disentangled spectra to measure the RVs is bound to reproduce the same biased RVs with which the disentangled spectra were obtained. }

{Secondly, the \cite{Gonzalez2006} algorithm has been found to be sensitive to local minima in the disentangling parameter space. This has arguably led to erroneous SB2 solutions \citep[e.g. 9~Sgr,][]{2012A&A...542A..95R,2016A&A...589A.121R}, which were identified and corrected by \cite{2021A&A...651A.119F} using a ‘grid’- disentangling approach (such as we employ here).}
 
{In other words, we believe that the faulty analysis of previous approaches \citep[and specifically that of][]{Linder2008} hinges on the first measurement of the RVs, which was flawed, possibly coupled with a limitation of the disentangling analysis used.
}

{
In contrast, in this study, we do not measure the RVs {a priori}, but simply assume that both components exhibit anti-phased motion over the orbital period. Disentangling of multiple lines consistently yields extreme mass ratios (Results, subsection ``Spectral disentangling), which agrees with our independent orbital analysis using the isolated O\,{\sc iii}\,$\lambda 5592$ line. }

\section*{{Data Availability}}

{The spectra analysed during the current study are available in reduced form from the Canadian Astronomy Data Centre (CADC) CFHT archive,} 

The ESPaDOnS spectropolarimetric observations of Plaskett's star are publicly available from the Canadian Astronomy Data Centre (CADC) CFHT archive, under the spectrum identifiers listed in the radial-velocity table [\url{https://www.cadc-ccda.hia-iha.nrc-cnrc.gc.ca/en/cfht}]. The IUE spectrum is available from the MAST archive [\url{https://archive.stsci.edu}]. Photometry used in the SED fit (Johnson UBV, 2MASS JHKs) is taken from the literature as cited in the Methods. The radial velocities (63 spectra; Julian dates, velocities and uncertainties for both components), the disentangled spectra of both components (K1 = 204.67 km/s, K2 = 29.6 km/s) and the PoWR model spectra used for the fits (custom and grid models) are available at https://doi.org/10.5281/zenodo.24371501. Source data are provided with this paper for Figs. 1, 2, 4 and 6–11.

{All files required to reproduce our \texttt{MESA} simulations are available at} [\url{https://doi.org/10.5281/zenodo.18242245}].

\section*{{Code Availability}}
The spectral disentangling code is publicly available on GitHub at \url{https://github.com/TomerShenar/Disentangling_Shift_And_Add}. The cross-correlation code used for the radial-velocity measurements is publicly available on GitHub at \url{https://github.com/TomerShenar/Cross-correlation}. The scripts used for the orbital solution, the spectral and SED fits and the IUE P-Cygni analysis are available at \url{https://doi.org/10.5281/zenodo.24371501}.

\bibliography{sn-article}


\section*{Acknowledgments}
This research has made use of the SIMBAD database operated at CDS, Strasbourg (France), NASA's Astrophysics Data System (ADS) and the Canadian Astronomy Data Centre (CADC). 

{Based on observations obtained at the Canada-France-Hawai'i Telescope (CFHT) which is operated by the National Research Council of Canada, the Institut National des Sciences de l'Univers of the Centre National de la Recherche Scientifique of France, and the University of Hawai'i. CFHT is located on Maunakea on Hawai'i Island, a mountain of considerable cultural, natural, and ecological significance. Maunakea is a sacred site to Native Hawaiians, also known as Kānaka 'Oiwi. We would like to thank the Canada-France-Hawai'i Telescope (CFHT) Operations and Software Groups for their contributions and diligence in maintaining observatory operations; the CFHT Astronomy Group for their observation coordination and data acquisition efforts; and the CFHT Finance \& Administration Group for their contributions to the management and administration of the observatory. This work has made use of the julia programming language (https://github.com/JuliaLang/julia/blob/master/CITATION.bib), and the makie plotting library (https://scixplorer.org/abs/2021JOSS....6.3349D/abstract).}

\section*{Funding}

GAW acknowledges Discovery Grant support from the Natural Sciences and Engineering Research Council (NSERC) of Canada.

MAM acknowledges support from the ``La Caixa'' Foundation (ID 100010434) under the fellowship code LCF/BQ/PI23/11970035.

The Center for Computational Astrophysics at the Flatiron Institute is supported by the Simons Foundation.

TS acknowledges support from the Israel Science Foundation (ISF) under grant number 0603225041 and from the European Research Council (ERC) under the European Union's Horizon 2020 research and innovation program (grant agreement 101164755/METAL.

PM acknowledges support from the FWO senior fellowship number 12ZY523N and the European Research Council (ERC) under the European Union’s Horizon 2020 research and innovation program (grant agreement 101165213/Star-Grasp).

OK acknowledges support from the Swedish Research Council (grant number 2023-03667) and the Swedish National Space Agency.

\section*{Author contributions}

\begin{itemize}
\item GAW: Conceptualized the research project, conceptualized and led writing of the article, acquired spectroscopic data, analyzed RV measurements, contributed to discussion of magnetic field generation.
\item TS: Contributed to writing the article, performed RV measurements, performed spectroscopic analysis.
\item PM: Contributed to writing the article, performed MESA evolutionary modeling.
\item MAM: Contributed to writing the article, performed photometric analysis.
\item MC: Contributed to writing the article, developed theoretical estimates related to magnetic field generation.
\item JG: Contributed to writing the article, conceptualized the research project, performed initial exploration of RV measurements.
\item OK: Contributed to writing the article, performed initial exploration of RV measurements.
\item HS: Contributed to writing the article, performed initial exploration of spectra and initial spectroscopic analysis.
\item SdM: Contributed to writing the article, performed initial exploration of evolutionary scenarios. 
\end{itemize}

\section*{Competing interests}

The authors declare no competing interests.

\section*{Tables}

\section*{Figures}

\end{document}